\documentclass[twocolumn]{aastex631}

\usepackage[T1]{fontenc}

\DeclareRobustCommand{\VAN}[3]{#2}
\let\VANthebibliography\thebibliography
\def\thebibliography{\DeclareRobustCommand{\VAN}[3]{##3}\VANthebibliography}

\hypersetup{linkcolor=blue,citecolor=blue,filecolor=cyan,urlcolor=magenta}
\usepackage{graphicx}   
\usepackage{amsmath}    
\usepackage{amssymb}    
\usepackage{newtxtext,newtxmath}
\usepackage{color}
\usepackage{threeparttable}
\usepackage{hyperref}
\usepackage{ulem}

\def\d{\mathrm{d}}

\newcommand{\lta}{\lower 2pt \hbox{$\, \buildrel {\scriptstyle <}\over {\scriptstyle \sim}\,$}}
\newcommand{\gta}{\lower 2pt \hbox{$\, \buildrel {\scriptstyle >}\over {\scriptstyle \sim}\,$}}
\definecolor{blazeorange}{rgb}{1.0, 0.4, 0.0}
\definecolor{seagreen}{rgb}{0.18, 0.55, 0.34}
\definecolor{rufous}{rgb}{0.66, 0.11, 0.03}
\definecolor{royalfuchsia}{rgb}{0.79, 0.17, 0.57}
\definecolor{scarlet}{rgb}{1.0, 0.13, 0.0}
\definecolor{royalpurple}{rgb}{0.47, 0.32, 0.66}

\begin{document}
	
	\title{Ultrafast compact binary mergers}

	\correspondingauthor{Paz Beniamini}
	\email{pazb@openu.ac.il}
	
	\author[0000-0001-7833-1043]{Paz Beniamini}
	\affiliation{Department of Natural Sciences, The Open University of Israel, P.O Box 808, Ra'anana 4353701, Israel}
	\affiliation{Astrophysics Research Center of the Open university (ARCO), The Open University of Israel, P.O Box 808, Ra'anana 4353701, Israel}
	\affiliation{Department of Physics, The George Washington University, 725 21st Street NW, Washington, DC 20052, USA}
	
	\author[0000-0002-7964-5420]{Tsvi Piran}
	\affiliation{Racah Institute for Physics, The Hebrew University, Jerusalem, 91904, Israel}

	\begin{abstract}
		The duration of orbital decay induced by gravitational waves (GWs) is often the bottleneck of the evolutionary phases going from star formation to a merger. We show here that kicks imparted to the newly born compact object during the second collapse generically result in a GW merger time distribution behaving like  $dN/d\log t \propto   t^{2/7}$ at short durations, leading to  ultrafast mergers. Namely, a nonnegligible fraction of neutron star binaries, formed in this way, will merge on a time scale as short as 10 Myr, and a small fraction will merge even on a time scale less than 10 kyr. The results can be applied to different types of compact binaries. We discuss here the implications for binary neutron star mergers. These include unique short gamma-ray bursts (GRBs), eccentric and misaligned mergers, $r$-process enrichment in the very early Universe and in highly star-forming regions, and possible radio precursors. Interestingly, we conclude that among the few hundred short GRBs detected so far, a few must have formed via this ultrafast channel.
	\end{abstract}
	
	\keywords{Gravitational Waves(678) --- Compact binary stars(283) --- Neutron stars(1108) --- Core-collapse supernovae(304) --- Chemical abundances(224) --- Gamma Ray Bursts(629)}

	\section{Introduction}
	\label{sec:intro}
	The mergers of compact stellar binaries release a huge amount of energy in the form of gravitational waves (GWs). Such mergers are now routinely detected by GW interferometers such as LIGO and VIRGO \citep{LIGOcat3}. Other planned and proposed missions will enable the detection of these binary mergers at much earlier stages before the merger and up to greater distances, enabling a much more thorough investigation of the merging binaries' properties. In case one or both of the compact objects are neutron stars (NSs), their merger can also result in an extremely bright electromagnetic display, spanning from the associated gamma-ray bursts (GRBs) to $r$-process powered kilonovae and the afterglow emission generated as the blast waves powering these events crash into the external medium (see review by \citealt{Nakar2020} and references therein).

	A compact stellar binary follows a slowly shrinking spiral orbit owing to GW radiation. In particular, a binary neutron star (BNS) is often expected to take hundreds of millions of years or more to merge \citep{Piran1992,Zwart1998,Belczynski2018}. This impacts both the merger event itself and its location within the host galaxy. The expectations for the merger delays and locations depend on the orbital parameters of the binary just after the final (second) gravitational collapse and the  formation of the second NS. Those parameters depend critically on the interplay between the orbital velocity of the collapsing star, $v_{\rm kep}$ and the kick velocity, $v_{\rm kick}$ given to it during the collapse \citep{Blaauw1961,Kalogera1996,Fryer1997,Pfahl2002}.
	
	Various approaches have been explored in the literature for inferring the merger delays, the host galaxies, and offsets relative to the galaxy centers. Perhaps the most common approach has been to use a population synthesis analysis that considers the distributions of the properties of the systems that evolve to become BNSs and feeds those into a detailed Monte Carlo model of the binary's evolution to predict distributions of interest \citep[e.g.,][]{Tutukov1993,Zwart1998,Belczynski1999,Voss2003,Oslowski2011,Chruslinska2018,Vigna-Gomez2018}. While this approach is potentially very powerful, it also involves various uncertainties, due to the extent to which the properties of the progenitors can be directly constrained from observations and, most importantly, due to the complex physics involved in the binaries' evolution (e.g. during the common envelope phase), which poses significant theoretical and numerical challenges. An independent approach relies on using the observed properties of known binary pulsars, to put direct limits on the allowed delay time distribution \citep{BP2019,Andrews2019}. This method too, is limited by the relatively small ($\sim 20$) number of observed systems and possible biases in the sample. 
	A third way of constraining BNS delays is by considering the outcome of their mergers. This has been done using data of $r$-process stellar abundances \citep[e.g.,][]{Matteucci2014,Wehmeyer2015,Cote2016,Komiya2016,Simonetti2019}, 
	redshifts, galaxy types, and offsets from galaxy centers of short GRBs \citep[e.g.,][]{Guetta2005,Leibler2010,Fong2010,Dietz2011,WP2015,Ghirlanda2016,Zevin2022} and, most recently, using the GW-detected BNSs, GW170817 and GW190425 \citep[e.g.,][]{Blanchard2017,Levan2017,Pan2017,Romero-Shaw2020}. These methods, too, have their own, independent biases. For example, low-redshift host galaxies of short GRBs are more likely to be determined  than high-redshift ones \citep[see][for details]{O'Connor2022}. 
	
	While these different approaches have resulted in varying conclusions, the majority of these studies find that, to a first approximation, the delay time distribution can be fit as a power law with $dN/dt(t>t_{\rm min})\propto t^{-\alpha}$ with $1\lesssim \alpha \lesssim 2$ and $20\mbox{ Myr}\lesssim t_{\rm min}\lesssim 200\mbox{ Myr}$. An exception is the work of \cite{BP2019} who find a significant excess of binaries with merger times in the range $30\mbox {Myr}\lesssim t\lesssim 1$\,Gyr, as compared with an extrapolation of the delay time distribution from longer merger times. In this new work, we show that while most have merger times of order tens to hundreds of Myr there should exist a shallow ($t^{-\alpha_0}$ with $\alpha_0=5/7$) power-law tail of ultrafast mergers at $t<t_{\rm min}$. These systems are a minority of the population, and as such it is not surprising that their existence has been largely overlooked. Moreover, because of their short merger times, there is a strong bias against detecting them. Nonetheless, the prediction of their existence is robust, their implications are unique, and their direct identification could inform our understanding of the formation of BNSs.

	We develop a formalism for studying the effects of kicks on binaries during stellar collapse. We consider collapses involving variable amounts of mass ejection, imparting a velocity kick on the collapsing star in some arbitrary direction. When the kick is comparable in magnitude and with opposite orientation to the Keplerian velocity, the collapsing star almost completely stops in its tracks. The angular momentum of the binary reduces significantly relative to the center of mass (CM); the binary becomes extremely eccentric, and consequently, it merges on an ultrafast timescale.
	Generalizing to a population of objects, if the fraction of systems with kick velocities comparable to the Keplerian velocity is significant, and if the kick orientation is drawn from an isotropic distribution in the collapsing star's frame, then the fraction of such ultrafast mergers can be remarkable, typically of order a few percent. Previous works have realized the importance of supernova (SN) kicks in leading to fast mergers \citep{Belczynski2002,O'Shaughnessy2008,Michaely2018}. Here we focus on the exact delay time distribution that arises from the kicked binaries and its correlation with other observables. We find that a {\it shallow} and {\it generic} power-law GW merger delay time tail develops, going down to arbitrarily short merger times.
	These systems are highly eccentric (with a potentially nonnegligible eccentricity even as they enter GW detectors' frequency range) and the collapsed star's spin is often significantly misaligned with the final binary orbit. 
	
	The formalism can be applied to different binaries that form by stellar collapse composed of a star that collapses to an  NS or black hole (BH) and a main-sequence, white dwarf (WD), NS or BH companion. 
	Indeed, it can even be extended to other types of kicks received by binary systems, such as the relatively weaker natal kicks associated with the formation of WD-WD binaries \citep{Hamers2019} or kicks received by wide-orbit binaries owing to flybys.
	Here we pay particular attention to BNS systems and show that the conditions for the development of the ultrafast merger tail hold for their progenitors. The result is the development of an ultrafast merger tail as described above. Previous studies provided evidence for a ``fast" population of BNS mergers having typical merger times of tens of Myrs \citep{Belczynski2006,Tauris2013,BP2019} with implications regarding the merger environments/host galaxy offsets \citep{Tsujimoto2014,ji2016Nature,BHP2016b,PB2021}, $r$-process enrichment \citep{HBP2018,Cote2019ApJ,Simonetti2019} and the brightness of the associated GRB afterglows \citep{Duque2020}.
	Here we show that $\sim 1\%$ of BNSs will be ``ultrafast" with shorter merger times than for the ``fast" channel and with a significant tail of the delay time distribution going down to order-of-magnitude-shorter time delays.
	
	The paper is organized is follows. In \S \ref{sec:mergerdelay} we present the basic equations for the orbital parameters, discuss some generic limits on their post-collapse values, and calculate the disruption probability. In \S \ref{sec:Mergerdelay} we present the change in GW merger times due to the collapse and develop our main formalism for calculating the delay time distribution for given mass ejection and kick, with an arbitrary orientation for the latter. In \S \ref{sec:aandeandalpha} we generalize the results to a situation in which there is a distribution of kick and/or Keplerian velocities and also discuss the resulting distributions of the separation, eccentricity, and misalignment degree. In \S \ref{sec:paramspace} we turn specifically to BNS systems, review several observed properties, and use them to infer the approximate conditions during the second stellar collapse. Then, in \S \ref{sec:bns} we discuss the implication regarding the BNS delay time distribution, observable imprints on GW detections, $r$-process enrichment, short GRBs, and radio precursors of BNS mergers. Readers that are mostly interested in the implications for BNS mergers can go directly to these last two sections.
	We conclude in \S \ref{sec:summary}.
	
	\section{Orbital and velocity changes due to  collapse, mass ejection, and kick - overview and general expression}
	\label{sec:mergerdelay}
	Consider a binary system composed of an NS and a companion star. The binary has an initial total mass $M_i$ and a separation $a_i$. Due to common envelope evolution \citep{Paczynski1976}, we can, to a good approximation, assume that the system is in a circular orbit before the collapse of the companion\footnote{Note, however, that population synthesis studies suggest that a sizeable fraction of binaries might be eccentric as they enter the common envelope phase \citep{Vigna-Gomez2020} and recent common envelope modelling suggests that a fraction of these might still maintain some residual eccentricity post-common envelope \citep{Trani2022}. We stress that while the assumption of $e_i=0$ in the present work significantly simplifies the expressions and provides clarity, it is by no means required, and that the most important results of the present work (such as the ultrafast merger time tail) are independent of this assumption.}. The collapse of the companion results in its envelope being ejected away from the system at a large velocity. This results in a sudden (relative to the orbital period, which is the typical time-scale in the system) mass loss from the system\footnote{The typical Keplerian velocities of the BNS systems are from tens to hundreds of hundreds km/s (see \S \ref{sec:bns}). This should be contrasted with the velocity of the ejecta, which is typically $> 3000$km./s \citep[see e.g.][]{Janka2022}. These ejecta velocities are much larger than the ``linearized" shell velocity (i.e. the mass-weighted mean velocity after averaging over the velocities of the quasi-isotropic ejecta, and which is equal to $v_{\rm kick} M_{2f}/[(\chi-1)M_f]$). Therefore, the assumption that the ejecta is removed instantaneously is a good approximation for BNS mergers. Furthermore, most of the following discussion is general and goes beyond the specific application to BNSs. There are a wide range of other applications (e.g. formations of NS-MS, BH-MS, BH-BH etc.) for which this assumption holds very well.} and imparts momentum to the remaining binary. If the mass ejection is nonspherical, the collapsing star may be given a velocity $\vec{v}_{\rm kick}$ relative to its orbital velocity (the mean velocity of the ejected envelope is larger by a factor of the mass ratio between the ejecta mass and the collapsing star).
	
	\subsection{Bound systems}
\begin{figure}
		\centering
		\includegraphics[scale=0.3]{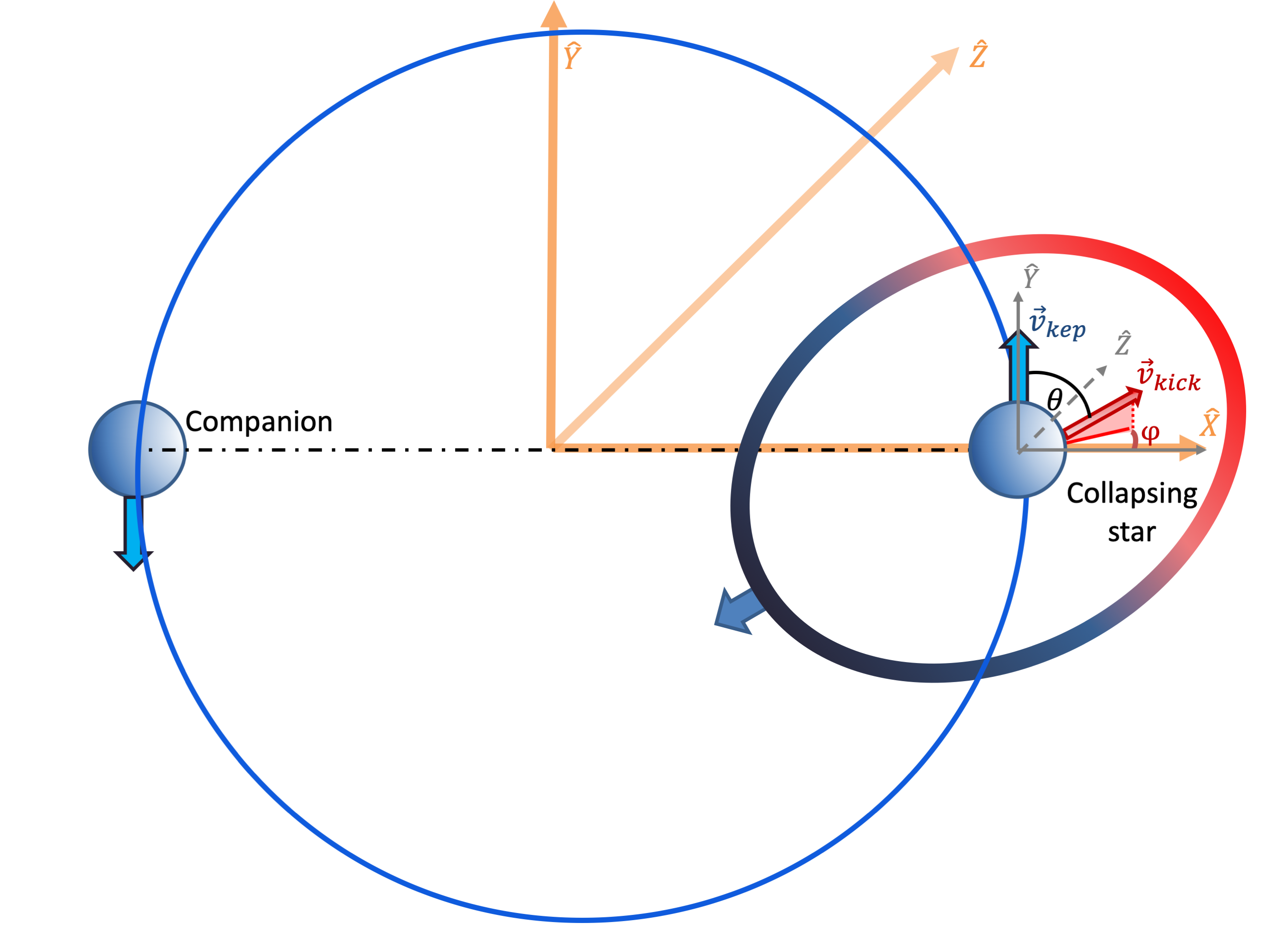}
		\caption{\small A schematic representation (not to scale) of the geometry at the moment of collapse.  Shown are the two stars and the SN shell. The latter could, in general, be asymmetric in terms of its matter and/or velocity distribution. In this situation, it carries with it nonzero linear momentum and imparts a kick, $\vec{v}_{\rm kick}$, on the collapsing star. $\theta$ is the angle between $\vec{v}_{\rm kep}$ (which is in the $\hat{Y}$ direction) and $\vec{v}_{\rm kick}$, and $\phi$ is the angle between projection of $\vec{v}_{\rm kick}$ on the plane perpendicular to $\vec{v}_{\rm kep}$ (the $\hat{X}-\hat{Z}$ plane) and the line connecting the collapsing star and its companion ($\hat{X}$). }
		\label{fig:schematic}
	\end{figure}

	If the system remains bound, the post-collapse orbital parameters are related to the pre-collapse parameters by (see, e.g. \citealt{Kalogera1996,Postnov2014}),
	\begin{equation}
		\frac{a_i}{a_f}=\bigg[ 2-\chi \bigg( \frac{v_{\rm kick}^2 \sin^2(\theta)+(v_{\rm kep}+v_{\rm kick} \cos(\theta))^2}{v_{\rm kep}^2}\bigg)\bigg] \ , 
	\end{equation}
	\begin{equation}
		1-e^2=\chi \frac{a_i}{a_f} \bigg( \frac{v_{\rm kick}^2 \sin^2(\theta)\sin^2(\phi)+(v_{\rm kep}+v_{\rm kick} \cos(\theta))^2}{v_{\rm kep}^2}\bigg) \ , 
	\end{equation}
	where $a_f$ is the final separation, $\chi=M_i/M_f\geq 1$ is the fractional change in the binary's total mass, $v_{\rm kep}=\sqrt{G M_i/a_i}$ is the initial Keplerian velocity (this is the initial orbital velocity of the reduced mass), $v_{\rm kick}$ is the kick velocity, $\theta$ is the angle between $\vec{v}_{\rm kep}$ and $\vec{v}_{\rm kick}$ and $\phi$ is the azimuthal angle. Fig. \ref{fig:schematic} provides a schematic depiction of this geometry.
	Defining $y\equiv |v_{\rm kick}/v_{\rm kep}|\geq 0$, $\mu\equiv \cos \theta$ we have
	\begin{equation}
		\label{eq:aiaf}
		\frac{a_i}{a_f}=2-\chi (1+2\mu y+y^2) \leq 2 \ , 
	\end{equation}
	\begin{eqnarray}
		\label{eq:esquared}
		&e^2=1\!-\!\chi [2\!-\!\chi (1\!+\!2\mu y\!+\!y^2)][(1\!+\!\mu y)^2\!-\!(\mu^2\!-\!1)y^2\sin^2\phi] \nonumber \\ & \leq 1-\chi [2-\chi(1+y)^2](1+y)^2 \ .
	\end{eqnarray}
	
	The equality sign on the RHS of Eq. \ref{eq:aiaf} corresponds to $\mu=-1, y=1$, i.e. a kick with equal magnitude to the Keplerian velocity and with opposite orientation. In such a situation, the relative velocity of the binary components cancels out exactly, and therefore the kinetic energy of the reduced mass of the system becomes zero, and only the gravitational component remains. Since the separation is unchanged at the moment of collapse, equipartition leads to the semi-major axis being reduced by at most a factor of 2. In the same limit, the eccentricity approaches unity (see Eq. \ref{eq:esquared}), and the components of the binary head toward each other on a collision course. It is in this regime of the phase space in which we expect ``ultrafast mergers", namely events that will merge on a time scale much shorter than the one corresponding to circular evolution from the initial separation.  
	
	It is useful to write the ratio of the final to initial binary angular momentum (per unit reduced mass) as a function of $a_i/a_f$ and $e$,
	\begin{equation}
		\label{eq:Jratio}
		\frac{j_f}{j_i}=\sqrt{ \chi^{-1} \frac{a_f}{a_i}(1-e^2)}<\sqrt{\frac{2}{\chi}} \ , 
	\end{equation}
	where we have used the fact that the new orbit must intersect with the old one, leading to $a_f(1-e)<a_i<a_f(1+e)$.
	
	The change of the binary's CM velocity is obtained from momentum conservation:
	\begin{eqnarray}
		\label{eq:delvcm}
		&\Delta \vec{v}_{\rm CM}=-\frac{\chi-1}{\chi}\frac{M_1}{M_f}\vec{v}_{\rm kep}+\frac{M_{2f}}{M_f}\vec{v}_{\rm kick}  \\
		& \Delta v_{\rm CM}=v_{\rm kep}\sqrt{\left(\frac{M_{2f}}{M_f}\right)^2y^2+\left(\frac{(\chi-1)M_1}{\chi M_f}\right)^2-\frac{2y(\chi-1)M_1M_{2f}}{\chi M_f^2}\mu} \nonumber 
	\end{eqnarray}
	where $M_1$ is the noncollapsing star and $M_{2f}$ is the final mass of the collapsing star.
	The first term in the top line of Eq. \ref{eq:delvcm} is known as the Blaauw kick \citep{Blaauw1961}. It represents the change in CM velocity that is obtained in the case where the mass ejection is completely symmetric in the exploding star's frame and it originates from the momentum carried by the ejecta shell relative to the pre-collapse CM frame. 
	Finally, for $M_{2f}=M_1=0.5M_f$ (typical for BNS formation) we see that $|\Delta v_{\rm cm}/v_{\rm kep}|\leq \frac{1}{2}(y+\frac{\chi-1}{\chi})$.

	\subsection{Disrupted binaries}
	\label{sec:disrupt}
	Non-disrupted binary solutions, in which we are interested, exist as long as $e<1$. We define the limiting angle for those solutions as $\mu_0\equiv \mu(e=1)$.
	Using Eq. \ref{eq:esquared} we find
	\begin{equation}
		\label{eq:mu0}
		\mu_0=\frac{2-\chi-\chi y^2}{2\chi y} \ .
	\end{equation}
	Requiring $-1\leq \mu_0 \leq1$ leads to
	\begin{equation}
		\label{eq:lim}
		\max(-1+\sqrt{\frac{2}{\chi}},1-\sqrt{\frac{2}{\chi}}) \leq y \leq 1+\sqrt{\frac{2}{\chi}} \ .
	\end{equation}
	We then calculate the disruption/survival probabilities, $P_{\rm dis}$ and $P_{\rm sur}$ respectively  \citep[see also][]{Sutantyo1978,Hills1983,Tauris2017},
	\begin{eqnarray}
		\label{eq:Psur}
		&	P_{\rm dis}=\int_{\mu_0}^{1}  d\mu / \int_{-1}^{1} d\mu=\frac{1-\mu_0}{2}=\frac{\chi(1+y)^2-2}{4\chi y}  \nonumber\\
		& P_{\rm sur}=1-P_{\rm dis}=\frac{1+\mu_0}{2}=\frac{2-\chi(1-y)^2}{4\chi y} \ . 
	\end{eqnarray}
	Notice that this expression only applies for values of $y$ within the limits given by Eq. \ref{eq:lim}. 
	In particular, (i) $\mu_0>1$ when $y<-1+\sqrt{\frac{2}{\chi}}$, leading to $P_{\rm dis}=0$. In this limit the maximal eccentricity (obtained when $\mu=1$), given by Eq. \ref{eq:esquared}, is less than unity. To first order in the kick and mass ejection, $e\lesssim (\chi-1)+2y$. (ii) $\mu_0<-1$ when $y<1-\sqrt{\frac{2}{\chi}}$ or $y>1+\sqrt{\frac{2}{\chi}}$ leading to disruption, $P_{\rm dis}=1$. Putting it all together, 
	\begin{equation}
		P_{\rm sur}\!= \!\!\left\{ \begin{array}{ll}1 & y<-1+\sqrt{\frac{2}{\chi}}\ ,\\
			\frac{2-\chi(1-y)^2}{4\chi y} & \max\left(-1+\sqrt{\frac{2}{\chi}},1-\sqrt{\frac{2}{\chi}}\right)<y<1+\sqrt{\frac{2}{\chi}}\ ,\\
			0 & y>1+\sqrt{\frac{2}{\chi}} ,\\
			0 & y<1-\sqrt{\frac{2}{\chi}}.
		\end{array} \right.
	\end{equation}
	The top panel of Fig. \ref{fig:chiyparamspace} shows the survival probability for general $\chi,y$.
	
	For $\chi>2$ the range of $y$ for which the system may survive the kick scales as $\Delta y\propto \chi^{-1/2}$ and is centered around $y=1$. Furthermore, for values of $y$ within that range, Eq.\ref{eq:Psur} shows that $P_{\rm sur}(y\approx 1)\approx (2\chi)^{-1}$. As a result, if $y$ is broadly distributed (as will be discussed in \S \ref{sec:ydist}), then (for $\chi>2$) the overall survival probability scales as $P_{\rm sur}(y\approx1)\cdot \Delta y\propto\chi^{-3/2}$.
	
	The velocity of the newly formed compact object is
	\begin{eqnarray}
		\label{eq:v2f}
		&\vec{v}_{\rm 2f}\!=\!\frac{M_1}{M_i}\vec{v}_{\rm kep}+\vec{v}_{\rm kick} \to \\ &  v_{\rm 2f}\!=\!v_{\rm kep}\sqrt{\left(\frac{M_1}{M_i}\right)^2\!+\!y^2\!+\!2y\mu \frac{M_1}{M_i}}\leq v_{\rm kep} \left(\frac{M_1}{M_i}+y\right). \nonumber
	\end{eqnarray}
	For $y\gtrsim M_1/M_i$, $v_{2f}$ is dominated by the kick velocity. Otherwise, it is a fraction of order $M_1/M_i$ times the Keplerian velocity.

	\section{Merger delays from kicks}
	\label{sec:Mergerdelay}
	The merger time of a binary with a semi-major axis $a$, eccentricity $e$, total mass $M=M_1+M_2$ and reduced mass $m=M_1 M_2/(M1+M_2)$ due to GW radiation can be approximated by
	\begin{equation}
		\label{eq:tmergerfull}
		t_{\rm m}\approx \frac{5a^4 (1-e^2)^{7/2}c^5}{256G^3 M^2 m}.
	\end{equation}
	We note that an explicit analytic formula does not exist for general $a, e$. The solution above is exact for $e\to 0$ and it underestimates the merger time by a modest factor of $\approx 1.8$ at the $e\to 1$ limit \citep{Peters1964}, while evolving monotonically between those values at intermediate values  (see e.g. figure 4.1.2 of \cite{Maggiore:2007ulw}  and an analytic fit by
	\citealt{Mandel2021}).
	
	Eq. \ref{eq:tmergerfull} shows that the merger time after the kick depends on 8 parameters: $a_i,M_i,m_i$ (determining the initial merger time for a circular orbit), $y,\chi,\mu,\phi$ (determining the change in orbit due to the kick) and on the change in the reduced mass of the binary $m_f/m_i\leq 1$. It is, therefore, constructive to consider the ratio of the post-collapse to pre-collapse merger time, which depends on only 5 parameters:
	\begin{equation}
		\label{eq:tratio}
		\frac{t_{\rm m,f}}{t_{\rm m,i}}\approx \left(\frac{a_i}{a_f}\right)^{-4}(1-e^2)^{7/2} \chi^{2} \frac{m_i}{m_f} \ . 
	\end{equation}
	Since the merger time ratio depends linearly on $\chi^{2} m_i/m_f$, it is useful to define $\tau$ such that
	\begin{eqnarray}
		\label{eq:tmerg}
		&\tau\equiv \left(\frac{a_i}{a_f}\right)^{-4}(1-e^2)^{7/2}= \\ & \left[2\!-\!\chi (1\!+\!2\mu y\!+\!y^2)\right]^{-1/2}\chi^{7/2} [(1+\mu y)^2\!-\!(\mu^2\!-\!1)y^2\sin^2\phi]^{7/2} \nonumber
		\\ & =\left[2\chi y (\mu_0-\mu)\right]^{-1/2}\chi^{7/2} [(1+\mu y)^2\!-\!(\mu^2\!-\!1)y^2\sin^2\phi]^{7/2}  \ .  \nonumber
	\end{eqnarray}
	The advantages of $\tau$ are as follows: (i) It is linearly proportional to the merger time ratio; (ii) It depends only on four rather than five parameters (in particular, this removes the dependence on the individual stellar masses in the binary);  (iii) It becomes identical to the delay time ratio when the ejected mass is small compared to the initial mass of the collapsing star (in which case $\chi^{2} m_i/m_f \approx 1$).
	Finally, it provides a useful limit. Since  $\chi^{2}{m_i}/{m_f}\geq 1$, it follows that the merger time ratio is longer than or equal to $\tau$: $t_{\rm m,f}/t_{\rm m,i}\geq \tau$.
	
	In the following, we analyze the distribution of $\tau$ resulting from different kicks and different amounts of mass ejection. Assuming that any directional asymmetry in the collapse is decoupled from (and therefore independent of) the collapsing star's orbital velocity, 
	$\mu, \phi$ are drawn from an isotropic distribution. Thus, the distribution of $\tau$, marginalized over kick orientations, is a function of only 2 parameters (or their distributions), $y, \chi$. 
	In \S \ref{sec:deltadelay} we consider fixed values of the latter and calculate the resulting probability distribution of $\tau$, $dP/d\tau$, for each pair of $y,\chi$ values. We then, in \S \ref{sec:ydist}, use those results to obtain $dP/d\tau$ in the more general case in which $y$ and/or $\chi$ are themselves drawn from some underlying distributions. We find that, quite generally,  a sizeable fraction of systems  obtain very low values of $\tau$. These  correspond to cases with $y\approx 1$ and $\mu \approx -1$, which, as mentioned in \S \ref{sec:mergerdelay}, lead to a large reduction in the orbital angular momentum and therefore to an extremely eccentric orbit, or a near head-on collision between the remaining binary components.

	\subsection{The delay time distribution for a constant kick magnitude and a constant mass ejection}
	\label{sec:deltadelay}
\begin{figure}
		\centering
		\includegraphics[scale=0.2]{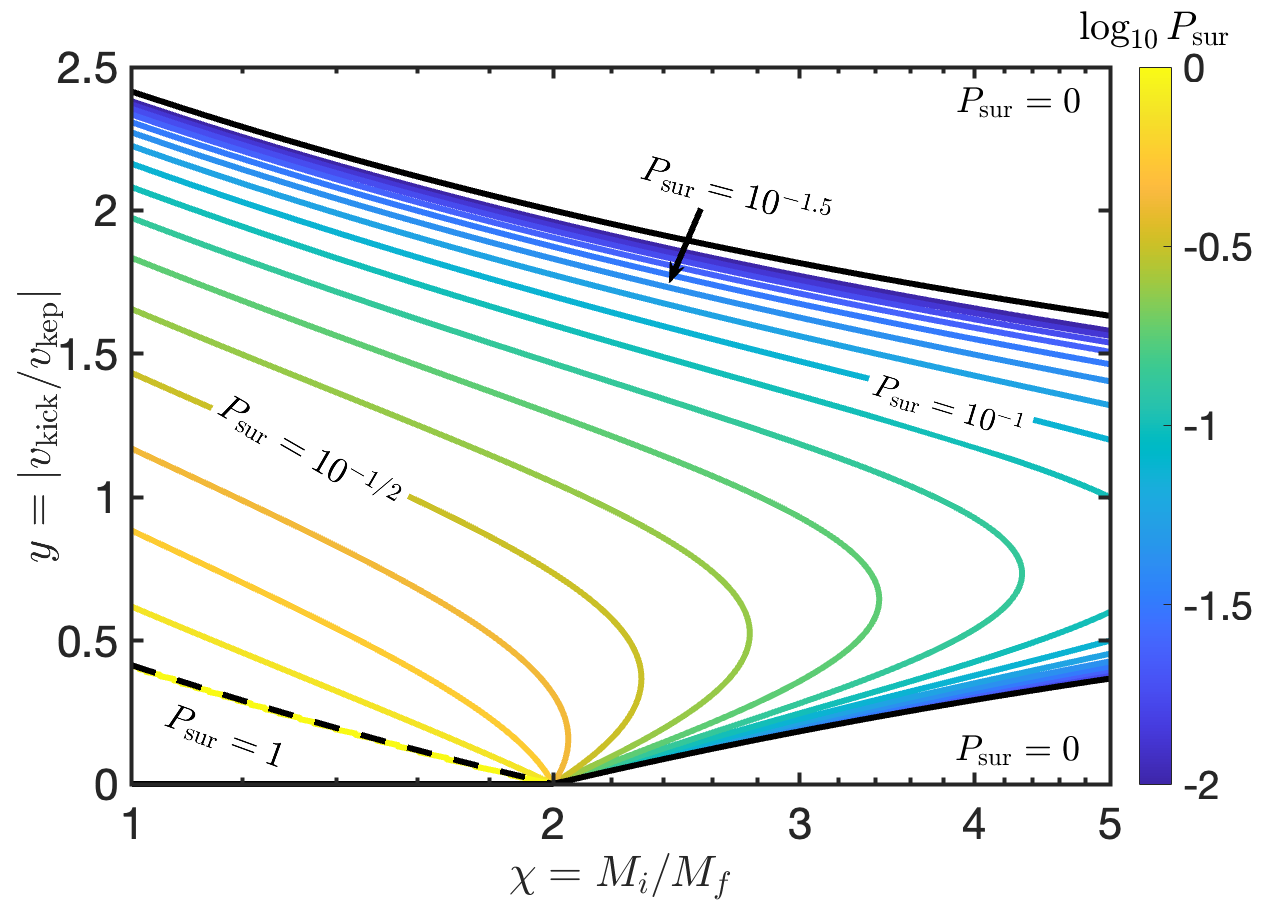}\\
		\includegraphics[scale=0.2]{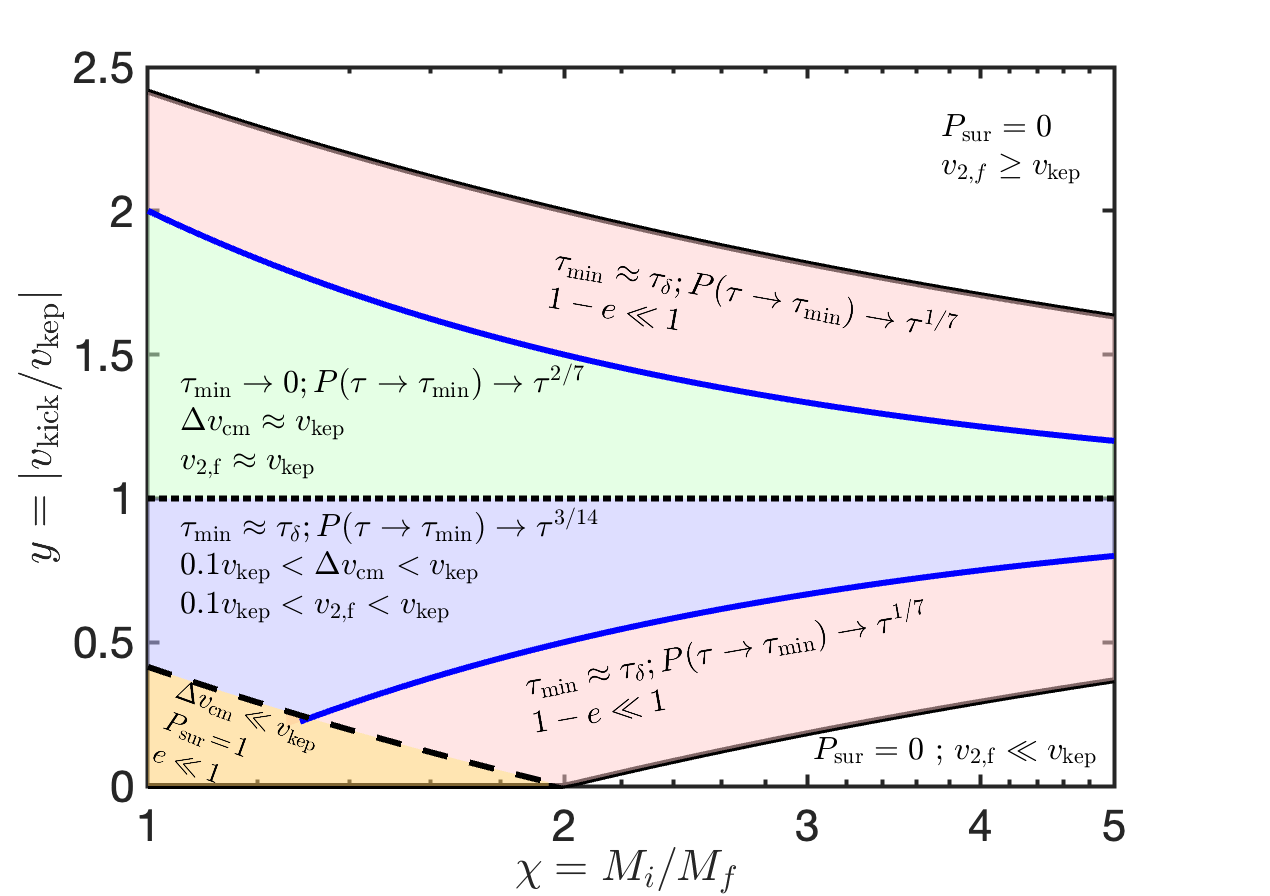}\\
		\includegraphics[scale=0.2]{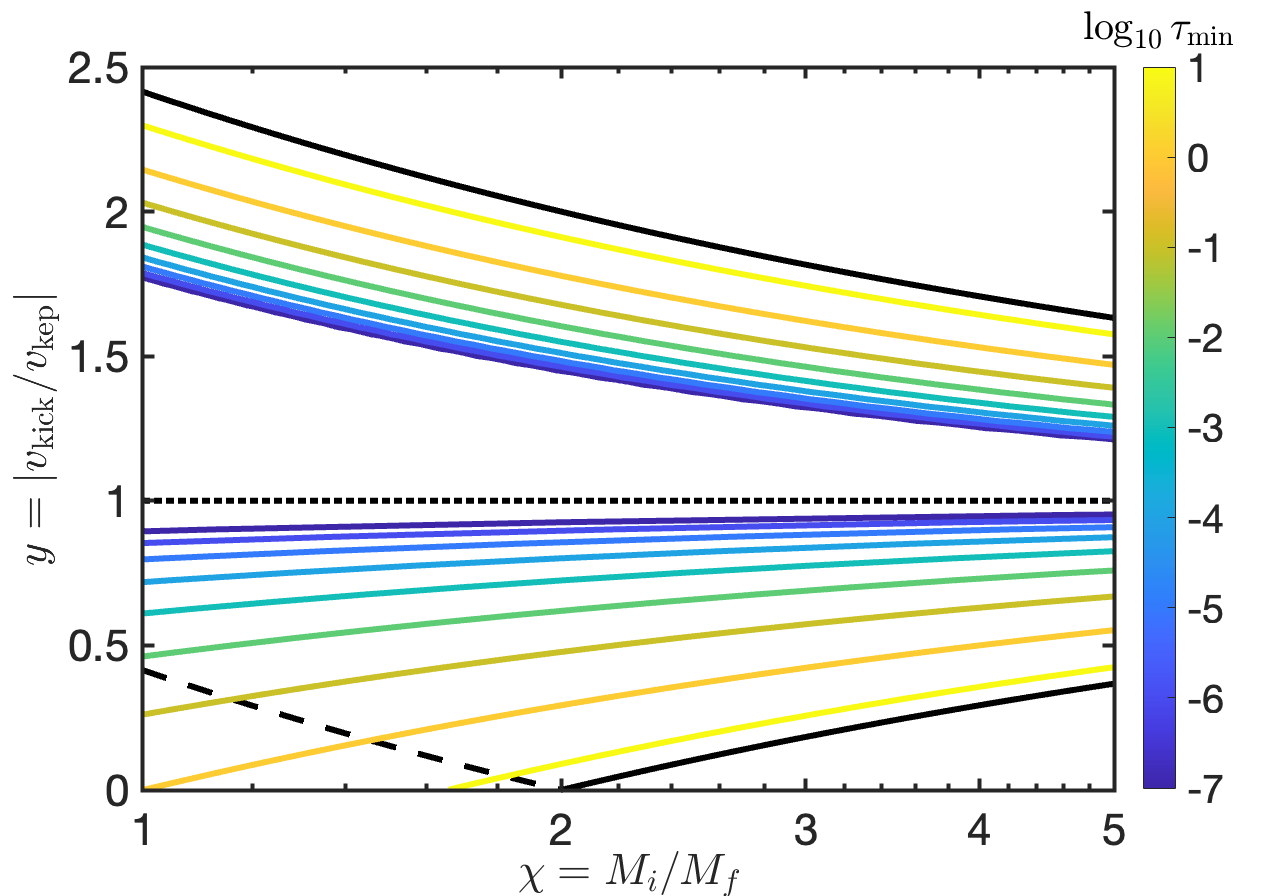}
		\caption{\small Top: The survival probability for different values of $\chi, y$. Black solid lines depict $y=1\pm\sqrt{2/\chi}$, beyond which the system is always disrupted. A dashed line depicts $y=-1+\sqrt{2/\chi}$, beyond which the binary always survives the kick.
			Middle: Types of delay time distributions that are obtained in this parameter space. Blue solid lines mark the transition $|y-1|=|\delta|=\chi^{-1}$ (corresponding to $\tau_\chi = \tau_\delta$) which separates the different  delay time distributions (see \S \ref{sec:deltadelay}). In each region we provide the expression for $\tau_{\rm min}$ -  the lowest value of $\tau$ that can be obtained from such kicks and the asymptotic scaling of the delay time distribution around $\tau_{\rm min} $. Bottom: The numerical values of $\tau_{\rm min}$ in the same parameter space. The value of $\tau_{\rm min}$ in the central white region is $<10^{-7}$ and quickly approaches 0 when $1\leq y \lesssim 1+\chi^{-1}$.}
		\label{fig:chiyparamspace}
	\end{figure}

	We consider first, in this section, the delay time ratios obtained for a fixed kick magnitude $y$ (or equivalently  $\delta$, where for $|y-1| \ll 1$ we define $\delta\equiv y-1$) and a fixed  mass ejection, $\chi$, varying isotropically the  kick orientations, $\mu$ (or $\mathscr{m}$ instead, where $ \mathscr{m} \equiv \mu -1$) and $\phi$ \footnote{The parameters $\delta, \mathscr{m}$ naturally become small around $y=1,\mu=-1$ which as mentioned is a critical point in terms of the delay time distribution behaviour. It is therefore instrumental to be able to expand various expressions as functions of small parameters in the vicinity of $y=1,\mu=-1$ as will be made explicit in the following discussion.}. We will present, next, analytic descriptions of the delay time 
	distribution that depend on the values of $y, \chi$ (or equivalently $\delta, \chi$). Fig.\ref{fig:chiyparamspace} depicts an overview of this parameter space and the distributions obtained within different regions.
	
	{\bf Rapid mergers:} 
	We define rapid mergers as those for which $\tau\ll 1$ (see Eqns. \ref{eq:tratio}, \ref{eq:tmerg}).
	From energy conservation,  the semi-major axis can decrease at most by a factor of 2 (as clear also from Eq. \ref{eq:aiaf} {and the subsequent discussion}). Therefore, $a_i/a_{f}<2$, and using Eq. \ref{eq:aiaf} the first term in the last two rows of Eq. \ref{eq:tmerg} is $>2^{-1/2}$. The term,  $\chi^{7/2}$, is always $>1$.  Thus, rapid mergers can take place only if the last term in Eq. \ref{eq:tmerg} is $\ll 1$, or equivalently $1-e^2\ll 1$ (see Eq. \ref{eq:tmergerfull} and \citealt{Peters1964}).  As shown in \S \ref{sec:disrupt}, the main effect of increasing $\chi$ is to increase the probability of disruption.
	We conclude that the condition for rapid mergers depends predominantly on $y$. Furthermore, we notice that $\tau=(a_f/a_i)^{1/2} \chi^{7/2}(j_f/j_i)^{7}$ (see the last line of Eq. \ref{eq:tmerg}). Writing the merger time in this way, it becomes clear that rapid mergers 
	require $j_f\ll j_i$ and $a_f\approx a_i$. From Eq. \ref{eq:tmerg} we see that these conditions can be attained simultaneously, since the former requires $\mathscr{m}\approx \delta, \sin \phi\approx 0$ and the latter $\mathscr{m}\ll \mu_0-1$.
	This corresponds to $e\to 1$ and therefore to $y\approx1, \mu\approx -1$ as discussed in \S \ref{sec:mergerdelay}.
	We can therefore take $0\leq \mathscr{m}\ll 1, |\delta|\ll 1$ for rapid mergers. Plugging into Eq. \ref{eq:tmerg} we see that for rapid mergers 
	\begin{equation}
		\label{eq:taumdelta}
		\tau\approx\chi^{7/2}[(\mathscr{m}-\delta)^2+2 \mathscr{m} \sin^2\phi]^{7/2}=\chi^{7/2}(j_f/j_i)^{7}\ .
	\end{equation}
	Eq. \ref{eq:taumdelta} shows that $\tau$ is bound in the range $\tau(\sin^2\phi\!=\!0)\!<\!\tau\!<\!\tau(\sin^2\phi\!=\!1)$. At the lower end $\tau$ is limited by $\tau(\sin^2\phi\!=\!0)\propto [\chi^{1/2} (\mathscr{m}\!-\!\delta)]^7$ and at the higher end it is limited by $\tau (\sin^2\phi\!=\!1)\propto [2\chi (\mathscr{m}\!-\!\delta)]^{7/2}$. Since isotropicity implies $dP(\mathscr{m})/d\mathscr{m}\!=\!const$ it follows that for $\sin^2\phi\!=\!0$ $dP(\tau)/d\log \tau\propto \tau ^{1/7}$, while for $\sin^2\phi\!=\!1$, $dP(\tau)/d\log \tau \propto \tau ^{2/7}$. 
	The result for general $\mathscr{m},\phi$ must be bound between these limits. 
	
\begin{figure*}
		\centering
		\includegraphics[scale=0.2]{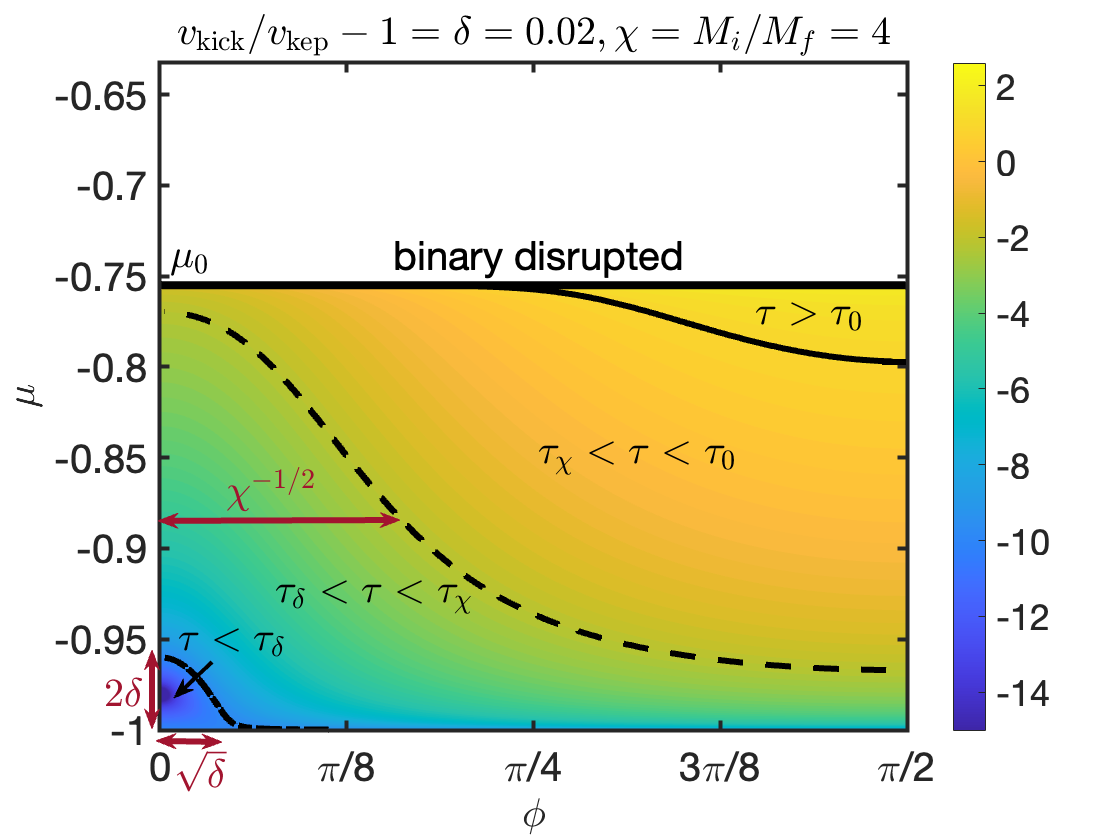}
		\includegraphics[scale=0.2]{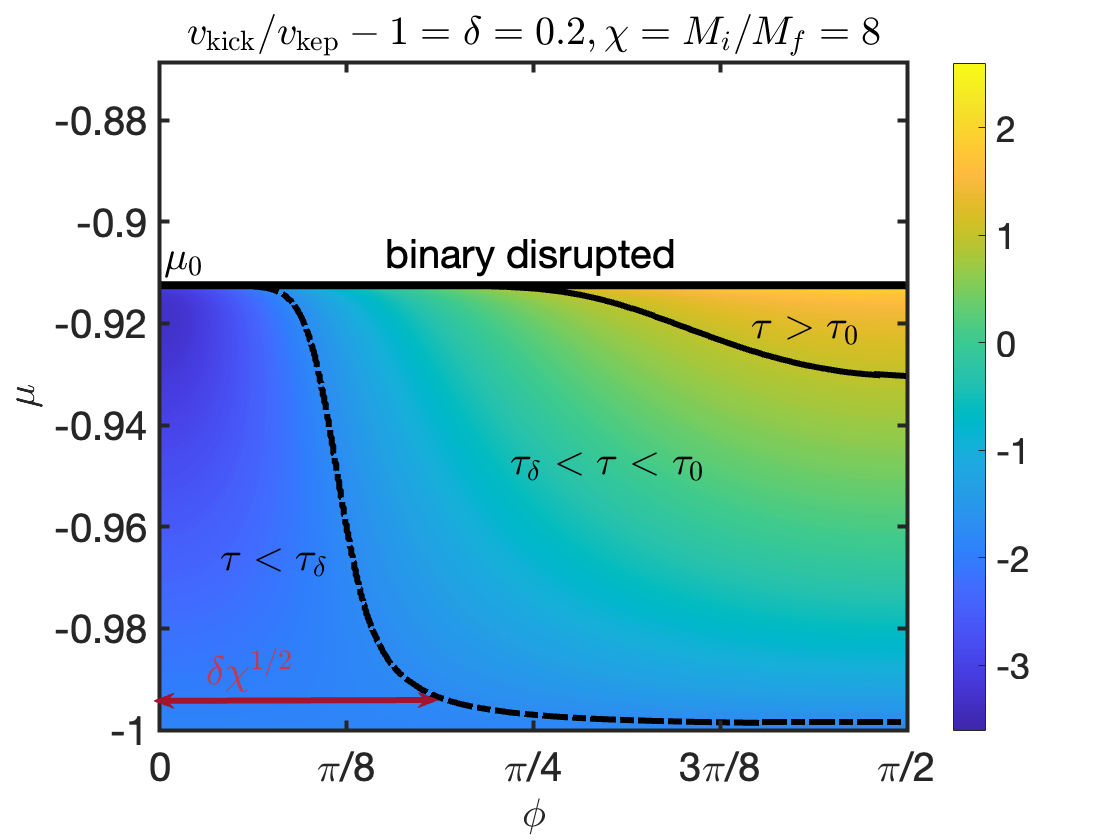}\\
		\includegraphics[scale=0.2]{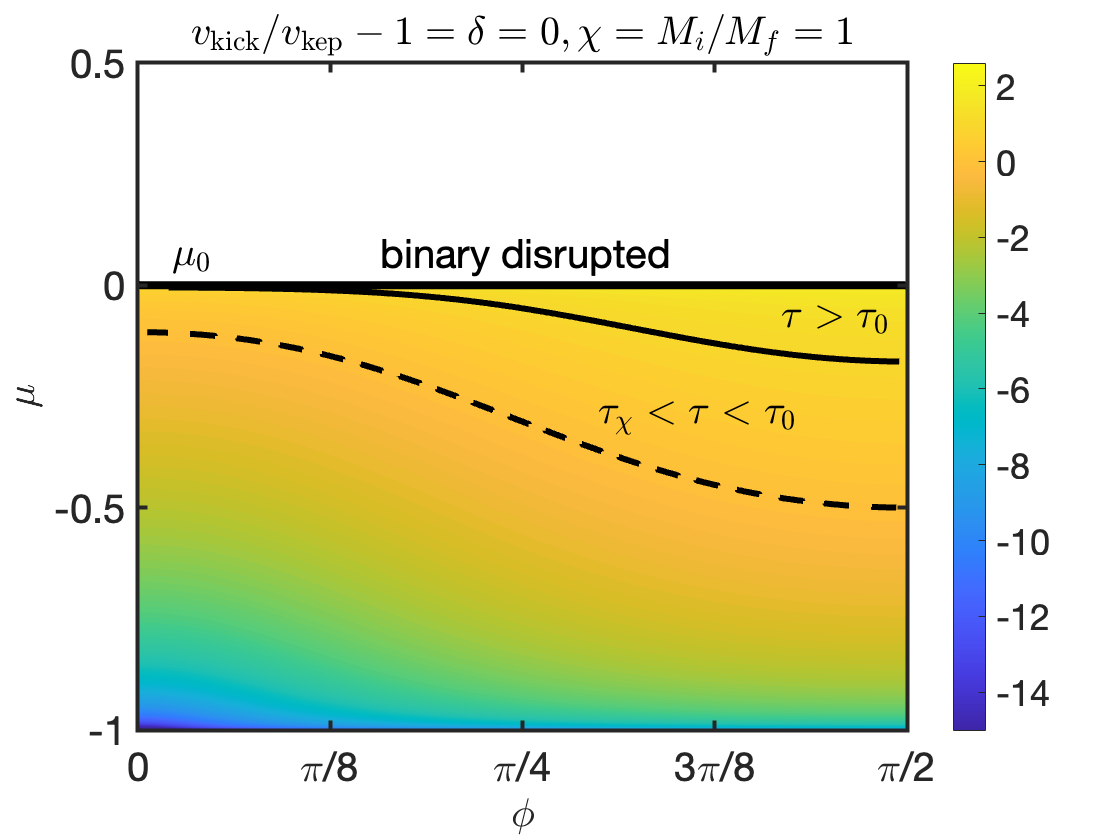}
		\includegraphics[scale=0.2]{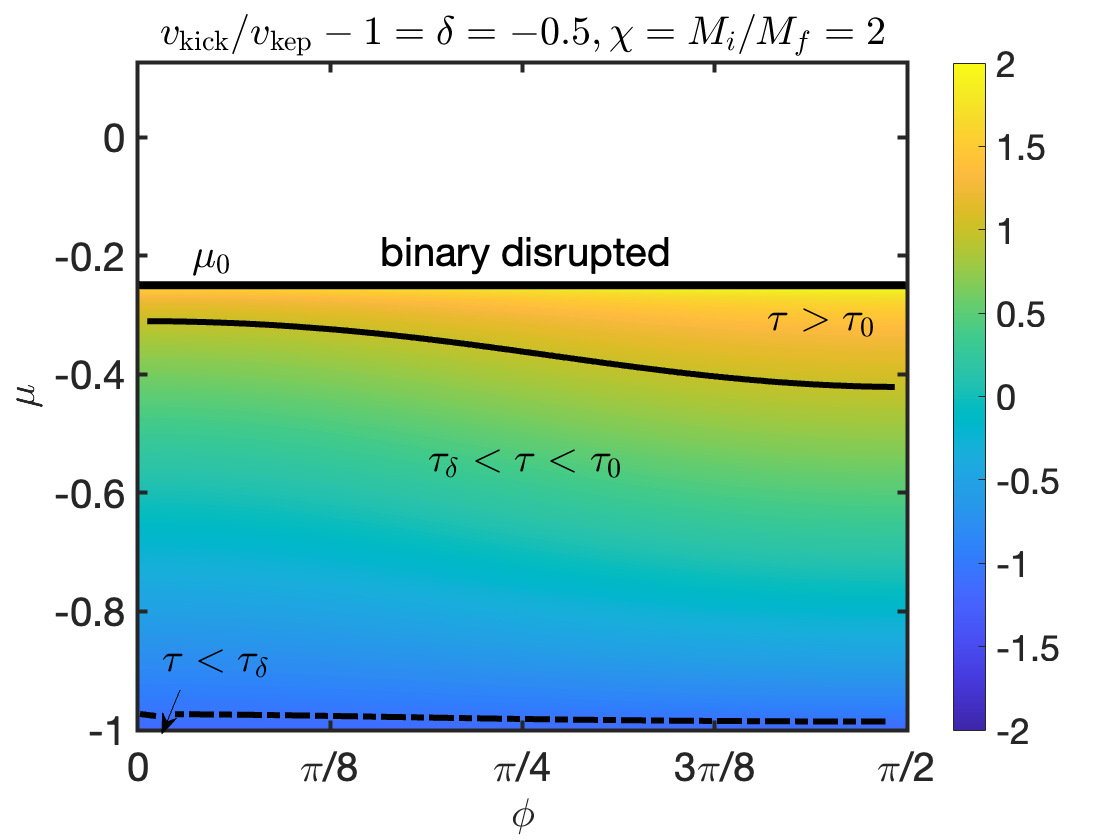}
		\caption{\small A color map describing the value of $\log_{10}(\tau)$ (i.e merger delay immediately post-collapse relative to that corresponding to the pre-collapse orbit) for different kick orientations (denoted by $\mu, \phi$) and for fixed values of $\chi=M_i/M_f$ and $y=v_{\rm kick}/v_{\rm kep}$ (or equivalently $\delta=y-1$). The different regimes of $\tau$ are separated by solid ($\tau_0$), dashed ($\tau_{\chi}$) and dot-dashed ($\tau_{\delta}$) lines. A horizontal solid line denotes the critical value of $\mu=\mu_0\approx1+\chi^{-1}$ above which the system is disrupted. We also denote, using double sided red arrows the approximate ranges in $\mu, \phi$ that correspond to the $\tau_{\delta}$ and $\tau_{\chi}$ lines. These reproduce well the analytic estimates described in \S \ref{sec:deltadelay}. Fast mergers ($\tau\lesssim 1$) occur when $\mu\lesssim \mu_0-1/(2\chi)$ or $\phi\lesssim \pi/4$. In general, the merger times get shorter as $\sin \phi \to 0$ and $\mu \to -1+ \min(\delta,\chi^{-1})$. }
		\label{fig:muphiparamspace}
	\end{figure*}

	From Eq. \ref{eq:taumdelta}, we see that the merger time is minimal for $\mathscr{m}\!\approx\! \min(\delta,\chi^{-1})$ and $\sin\phi\!=\!0$ (see also Fig. \ref{fig:muphiparamspace}). This suggests that the value of $\delta$ relative to $\chi^{-1}$ will determine the distribution of $\tau$. Indeed, we will show below that $dP/d\log \tau$ at $\tau\lesssim 1$ behaves as a broken power law, with the characteristic timescale ratios 
	\begin{eqnarray}
		\label{eq:taudelchidef}
		\tau_{\delta}&\equiv &|\delta|^7\chi^{7/2} \ , \\ \nonumber
		\tau_{\chi}&\equiv& \chi^{-7/2} \ .
	\end{eqnarray}
	The relative order of these two depends on the value of $\delta$ as compared to $\chi^{-1}$ as anticipated above. Notice also that these $\tau$ values correspond to $j_f/j_i=\delta$ and $j_f/j_i=\chi^{-1}$ respectively.
	
	We begin by considering the case of $0<\delta<\chi^{-1}<1$\footnote{Most of the following discussion holds also for $\delta<0$ such that $|\delta|<\chi^{-1}$. We will explicitly mention the difference in that case at the relevant point in the discussion.} that corresponds to the green region in Fig. \ref{fig:chiyparamspace}.
	In this case $\tau_{\delta}<\tau_{\chi}<1$. The probability distribution of $\tau$ is given by
	\begin{equation}
		\label{eq:J}
		P(\tau)=\frac{1}{4\pi}\int_{\phi_-(\tau)}^{\phi_+(\tau)} d\phi  \int_{\mathscr{m}_-(\phi,\tau)}^{\mathscr{m}+(\phi,\tau)} d\mathscr{m} 
	\end{equation}
	where $\mathscr{m}_-,\mathscr{m}_+,\phi_-,\phi_+$ are the limits of the region in the $[\mathscr{m},\phi]$ parameter space, within which $\tau'(\mathscr{m},\phi)<\tau$.
	The relation $\tau=\tau(\mathscr{m},\phi)$ can be inverted to connect $\mathscr{m}$ to $\phi$ for a given $\tau$, 
	\begin{eqnarray}
		\label{eq:mbar}
		&   (\mathscr{m}-\delta)^2+2(\mathscr{m}-\delta)\sin^2\phi+2\delta \sin^2\phi-\tau^{2/7}/\chi=0\to \nonumber \\
		&\mathscr{m}-\delta=\sin^2\phi\left[-1 +\sqrt{1+\frac{\tau^{2/7}\chi^{-1}-2\delta \sin^2\phi}{\sin^4\phi}}\right]. 
	\end{eqnarray}
	This relation has distinct behaviours depending on the value of $\tau^{2/7}\chi^{-1}$ relative to $\delta^2$. 
	This is the origin of the critical delay time defined above $\tau_{\delta}=\chi^{7/2}|\delta|^7$. 
	For $\tau \ll \tau_{\delta}$, there is an intermediate range of $\sin \phi$, $0< \tau^{2/7}/ \chi \delta \lesssim \sin^2 \phi\lesssim 2\delta<1$, for which there are no solutions to $\mathscr{m}$ through Eq. \ref{eq:mbar} \footnote{This point applies only to $\delta>0$. As can be seen from the expression for $\tau$, since $\mathscr{m}>0$, we see that for $\delta<0$ one always has $(\mathscr{m}-\delta)^2>|\delta|^2$ and therefore $\tau\gtrsim\tau_{\delta}$, see bottom right panel of Fig. \ref{fig:muphiparamspace}.}.  This is shown explicitly in the $\mu-\phi$ parameter space in Fig. \ref{fig:muphiparamspace}.
	Moreover, for $\phi\approx \sin \phi<\sqrt{\tau^{2/7}/\chi \delta}<\delta^{1/2}$ (which dominates the integral in Eq. \ref{eq:J}), we can approximate the term in the square root of Eq.\ref{eq:mbar} as $\tau^{2/7}/\chi\phi^4$ and we get $\mathscr{m}-\delta\approx \tau^{1/7}\chi^{-1/2}<\delta$. 
	Therefore, we have
	\begin{equation}
		\label{eq:Ptaulesstaudelta}
		P(<\tau)\sim\int_{0}^{\sqrt{\tau^{2/7}/\chi\delta}} d\phi  \int_{\delta-\sqrt{\tau^{2/7}/\chi}}^{\delta+\sqrt{\tau^{2/7}/\chi}} d\mathscr{m} \sim \tau^{2/7}\chi^{-1}\delta^{-1/2},
	\end{equation}
	or equivalently $dP/d\tau \propto \delta^{-1/2} \chi^{-1}\tau^{-5/7}$. In particular, for $\tau=\tau_{\delta}$, we see that the ranges of $\mathscr{m}, \phi$ that contribute to Eq. \ref{eq:Ptaulesstaudelta} are $\Delta \mathscr{m}\approx 2\delta, \Delta \phi=\delta^{1/2}$. This is demonstrated in Fig. \ref{fig:muphiparamspace}. It allows us to easily estimate the fraction of systems with $\tau\!<\!\tau_{\delta}$, $P(<\tau_{\delta})\!\sim\! \delta^{3/2}\!=\!\tau_{\delta}^{3/14} \tau_{\chi}^{3/14}$ ($P(<\tau_{\delta})/P_{\rm sur}\!\sim\! \delta^{3/2}/\chi\!=\!\tau_{\delta}^{3/14} \tau_{\chi}^{-1/14}$).
	
	Consider next $\tau_{\delta}<\tau<\tau_{\chi}\equiv \chi^{-7/2}$. In this case, there is a solution to $\mathscr{m}$ for any value of $\phi$ (see Fig. \ref{fig:muphiparamspace}). In particular, for $\phi<(\tau^{2/7}/\chi)^{1/4}$, the term in the square root of Eq.\ref{eq:mbar} can again be approximated by $\tau^{2/7}/\chi\phi^4<\chi^{-1/2}$  and we get $\mathscr{m}-\delta\approx \tau^{1/7}\chi^{-1/2}<\chi^{-1}$. Thus, we have 
	\begin{equation}
		P(<\tau)\sim\int_{0}^{(\tau^{2/7}/\chi)^{1/4}} d\phi  \int_{0}^{\sqrt{\tau^{2/7}/\chi}} d\mathscr{m} \sim (\tau^{2/7}/\chi)^{3/4}
	\end{equation}
	leading to $dP/d\tau \propto \chi^{-3/4}\tau^{-11/14}$. In particular, for $\tau=\tau_{\chi}$, we see that the ranges of $\mathscr{m}, \phi$ that contribute to Eq. \ref{eq:Ptaulesstaudelta} are $\Delta \mathscr{m}\approx \chi^{-1}, \Delta \phi=\chi^{-1/2}$ (see Fig. \ref{fig:muphiparamspace}). This allows us to estimate the fraction of systems with $\tau\!<\!\tau_{\chi}$, $P(<\tau_{\chi})\!\sim\! \chi^{-3/2}\!=\!\tau_{\chi}^{3/7}$ ($P(<\tau_{\chi})/P_{\rm sur}\!\sim\! \chi^{-1/2}\!=\! \tau_{\chi}^{1/7}$).
	
	Finally, consider the case in which $\tau_{\chi}\ll \tau<1$. 
	For the system to remain bound, we have an upper limit on $\mathscr{m}$ such that $\mathscr{m}\!<\!1+\mu_0\!\approx \!\chi^{-1}$ (see Eq. \ref{eq:mu0}).
	We also note that, by virtue of Eq. \ref{eq:mbar}, the integral in Eq. \ref{eq:J} is dominated by the largest value of $m$ for which the system is not disrupted. Therefore, the requirement $\tau\!>\!\tau_{\chi}$ leads to $\mathscr{m}\!\approx \!\chi^{-1}$ and correspondingly $\phi\!\sim \!\sin \phi \!<\! \tau^{1/7}$  (see Eq. \ref{eq:taumdelta} and Fig. \ref{fig:muphiparamspace}) and
	\begin{equation}
		\label{eq:tau1over7}
		P(<\tau)\sim\int_{0}^{\tau^{1/7}} d\phi  \int_{0}^{1/\chi} d\mathscr{m} \sim \tau^{1/7}\chi^{-1}.
	\end{equation}
	Therefore, $dP/d\tau \propto \tau^{-6/7}\chi^{-1}$.

	Consider next the case $1>\delta>\chi^{-1}$. This time we have $\tau_{\chi}<\tau_{\delta}<1$. For the  system to remain bound, $\mathscr{m}<\chi^{-1}<\delta$, which leads to $\tau^{2/7}>\chi (\mathscr{m}-\delta)^2\gtrsim \chi \delta^2$ and therefore $\tau\gtrsim \chi^{7/2}\delta^7>\tau_{\delta}$. These conditions correspond to the red regions in Fig. \ref{fig:chiyparamspace}.
	As a result, we see that $dP/d\tau$ drops sharply to zero at $\tau\ll \tau_{\delta}$\footnote{This is true when $\delta\gg \chi^{-1}$. When $\delta\approx \chi^{-1}$ the exact value of $\tau$ at which the probability goes to zero is somewhat smaller than $\tau_{\delta}$. For the sake of clarity we ignore this marginal case here.}. Therefore, mergers cannot be arbitrarily rapid in this situation, and fast mergers exist only in the regime $\tau_{\delta}<\tau<1$. In this limit, the derivation is the same as in eq. \ref{eq:tau1over7} and we again get $dP/d\tau \propto \tau^{-6/7}\chi^{-1}$. In particular, for $\tau=\tau_{\delta}$, we see that, the ranges of $\mathscr{m}, \phi$ that contribute to Eq. \ref{eq:Ptaulesstaudelta} are $\Delta \mathscr{m}\approx \chi^{-1}, \Delta \phi=\delta \chi^{1/2}$ (see Fig. \ref{fig:muphiparamspace}).

	{\bf Slow mergers.}
	As mentioned above, $\tau=(a_f/a_i)^{1/2} (j_f/j_i)^{7}\chi^{7/2}$. Since from Eq. \ref{eq:Jratio}, $(j_f/j_i)^2<2/\chi$, we see that $\tau\gg1$ can only be obtained if $a_f\gg a_i$, which requires systems that are almost disrupted by the kicks.
	We can then write $\tau \approx [2\chi (\mu_0-\mu)]^{-1/2}(j_f/j_i)^{7}\chi^{7/2}$, which shows that slow mergers require $\mu_0-\mu\ll \chi^{-1}$. At this limit, $\mathscr{m}\approx \mu_0+1\approx \chi^{-1}$ and, as follows from Eq.\ref{eq:taumdelta}, for $\sin^2 \phi\sim 1$, $(j_f/j_i)^7\chi^{7/2}\sim 1$. As a result, $dP/d\tau$ becomes independent of $\phi$ and it instead becomes a direct function of $\mathscr{m}$. We thus have $dP/d\tau=(dP/d\mathscr{m})(d\mathscr{m}/d\tau)=\chi^{-1}\tau^{-3}$.

	{\bf The overall distribution.}
	Combining the sub-cases discussed above, we obtain an expression for the overall probability distribution of $\tau$. 
	It turns out that the scaling developed for $\tau<1$ provides a good approximation up to $\tau\approx 10$. We therefore define $\tau_0=10$ and write the distribution as a broken power law function of $\tau$ in regimes defined by $\tau_{\chi},\tau_{\delta},\tau_0$:
	\begin{eqnarray}
		\label{eq:deltalesschi}
		& \mbox{For }0<\delta\ll \chi^{-1}\mbox{, }  \\ &
		\frac{dP}{d\log \tau}\!\approx \!\frac{1}{15\chi-14\chi^{1/2}\tau_0^{-1/7}} \!\left\{ \begin{array}{ll}\tau_{\chi}^{-{1 \over 14}}\tau_{\delta}^{-{1 \over 14}}\tau_0^{-{1 \over 7}}\tau^{2 \over 7} & \tau<\tau_{\delta}\ ,\\
			\tau_{\chi}^{-{1 \over 14}} \tau_0^{-{1 \over 7}}\tau^{3 \over 14} & \tau_{\delta}<\tau<\tau_{\chi}\ ,\\
			(\tau/\tau_0)^{1 \over 7} & \tau_{\chi}<\tau<\tau_0 , \\
			(\tau/\tau_0)^{-2} & \tau>\tau_0.
		\end{array} \right.\nonumber
	\end{eqnarray}
	
	\begin{eqnarray}
		\label{eq:deltalesschimin}
		& \mbox{For }\delta<0, |\delta|\ll \chi^{-1}\mbox{, } \\ &
		\frac{dP}{d\log \tau}\!\approx \!\frac{1}{15\chi-14\chi^{1/2}\tau_0^{-1/7}} \!\left\{ \begin{array}{ll}0 & \tau<\tau_{\delta}\ ,\\
			\tau_{\chi}^{-{1 \over 14}} \tau_0^{-{1 \over 7}}\tau^{3 \over 14} & \tau_{\delta}<\tau<\tau_{\chi}\ ,\\
			(\tau/\tau_0)^{1 \over 7} & \tau_{\chi}<\tau<\tau_0 , \\
			(\tau/\tau_0)^{-2} & \tau>\tau_0.
		\end{array} \right.\nonumber
	\end{eqnarray}
	
	\begin{eqnarray}
		\label{eq:deltagreatchi}
		&\mbox{For }|\delta|\gg \chi^{-1}\mbox{, }\\
		& \frac{dP}{d\log \tau}\!\approx \!\frac{1}{15\chi-14\delta \chi^{3/2}\tau_0^{-1/7}} \!\left\{ \begin{array}{ll}0 & \tau<\tau_{\delta}\ ,\\
			(\tau/\tau_0)^{1/7} & \tau_{\delta}<\tau<\tau_0\ ,\\
			(\tau/\tau_0)^{-2} & \tau>\tau_0.
		\end{array} \right.\nonumber
	\end{eqnarray}
	Here  the normalization ensures that $\int (dP/d\tau) d\tau=P_{\rm sur}\approx (2\chi)^{-1}$.
	We see that as long as $\tau_{\delta}\ll 1$ \footnote{If this condition is not satisfied the median is of order $\tau_{\delta}$ which is the smallest $\tau$ that doesn't correspond to a disruption of the binary.}, the median value of $\tau$ is of order unity, while the distribution can extend to values of $\tau$ that are both much greater and much smaller than this median value. Moreover, the shallow slope of $dP/d\log \tau$ below the peak (evolving as $\tau^{1/7}$ or $\tau^{3/14}$), implies that a significant fraction of mergers will have $\tau\ll 1$. This will have important implications as will be explored further in \S \ref{sec:bns}.

	A slightly more accurate description of the probability distribution can be given by smoothing the distribution around the peak. To do this we define
	\begin{equation}
		f=0.17\chi^{-1} \tau^{1/7}\left[1+\left(\frac{\tau}{14}\right)^{27/28}\right]^{-20/9}
	\end{equation}
	and 
	\begin{equation}
		g=0.21\min(\chi^{-3/4}\tau^{3/14},\chi^{-1}\delta^{-1/2}\tau^{2/7})
	\end{equation}
	such that to a good approximation
	\begin{equation}
		\label{eq:deltalesschi2}
		\mbox{For }0<\delta\ll \chi^{-1}\mbox{, } \quad
		\frac{dP}{d\log \tau}\!\approx \min(f(\tau),g(\tau)),
	\end{equation}
	
	\begin{equation}
		\label{eq:deltalesschimin2}
		\mbox{For }\delta<0, |\delta|\ll \chi^{-1}\mbox{, } \quad
		\frac{dP}{d\log \tau}\!\approx \min(f(\tau),g(\tau))\Theta(\tau-\tau_{\delta})
	\end{equation}
	and
	\begin{equation}
		\label{eq:deltagreatchi2}
		\mbox{For }\delta\gg \chi^{-1}\mbox{, } \quad
		\frac{dP}{d\log \tau}\!\approx f(\tau)\Theta(\tau-\tau_{\delta})
	\end{equation}
	where $\Theta(x)$ is the Heaviside function.
	
\begin{figure}
		\centering
		\includegraphics[scale=0.2]{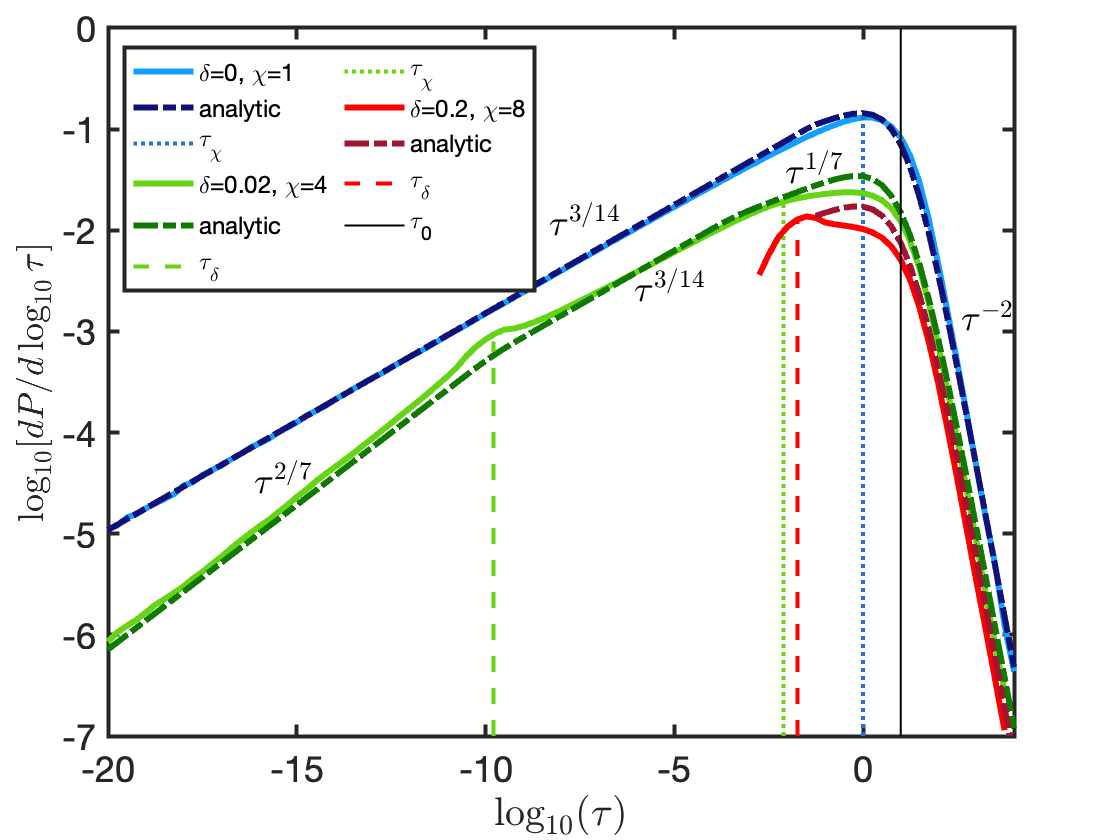}
		\caption{\small The delay time ratio ($\tau$) distribution resulting from kicks with different values of $\chi=M_i/M_f$ and $y=v_{\rm kick}/v_{\rm kep}$ (or equivalently $\delta\equiv y-1$) and with random orientations relative to the binary orbit. A solid colored line depicts the numerical solution to Eq. \ref{eq:tmerg} and the dashed line is our analytical fit described in Eqns. \ref{eq:deltalesschi2}, \ref{eq:deltagreatchi2}. For each case, a dotted line depicts $\tau_{\chi}$, a dashed line depicts $\tau_{\delta}$ and a solid black line depicts $\tau_0$. The red line, corresponding to $\delta = 0.2$ and $\chi=8$ ends abruptly at $\lesssim \! \tau_\delta$, where the systems disrupt. }
		\label{fig:comparedelta}
	\end{figure}

	The overall distribution of $\tau$ (averaged over all $\mu, \phi$) is shown in Fig. \ref{fig:comparedelta} in comparison with our analytic approximations. While the approximations provide a good description of the overall distribution (including normalization, all asymptotic behaviours and the cutoff values of $dP/d\tau$ when those exist), there are subtler features of the numerical distribution, such as the kink in $dP/d\tau$ around $\tau_{\delta}$, that are not captured by this analytic treatment. Eqns. \ref{eq:deltalesschi2}, \ref{eq:deltalesschimin2}, \ref{eq:deltagreatchi2} can be easily implemented in population synthesis codes and could provide a good description of the corresponding delay time distribution, while potentially saving important calculation time for codes focusing on other stages of the evolution.

	\subsection{The delay time distribution for widely distributed kick magnitude}
	\label{sec:ydist}
	
	The results of the previous section can be generalized to the case in which $y$ varies between events.
	Since rapid mergers correspond to a narrow region (of width $\Delta y\sim \chi^{-1}$, see Fig. \ref{fig:chiyparamspace}) around $y=1$, we can, to a good approximation, take the probability distribution, $dP/dy$ (or equivalently $dP/d\delta$), to be constant within that range.
	Using Eq. \ref{eq:deltalesschi} this leads to
	\begin{eqnarray}
		\label{eq:PLdistfast}
		& \frac{dP}{d\log \tau}\left(\tau\ll 1\right)\! \approx \!  \left.\frac{dP}{dy}\right \vert_{y=1}\int_0^{\chi^{-1}}\left.\frac{dP}{d\log \tau}\right \vert_{\delta, \tau \ll 1}d\delta \nonumber \\ &\approx \!\left.\frac{dP}{dy}\right \vert_{y=1}\frac{0.17}{\chi^{3/2}}\tau^{2/7}\approx P(0<\delta<\chi^{-1}) \frac{0.17}{\chi^{1/2}}\tau^{2/7}.
	\end{eqnarray}
	The distribution is proportional to $\tau^{2/7}$ and the normalization is proportional to $\chi^{-3/2}$.
	
\begin{figure*}
		\centering
		\includegraphics[scale=0.2]{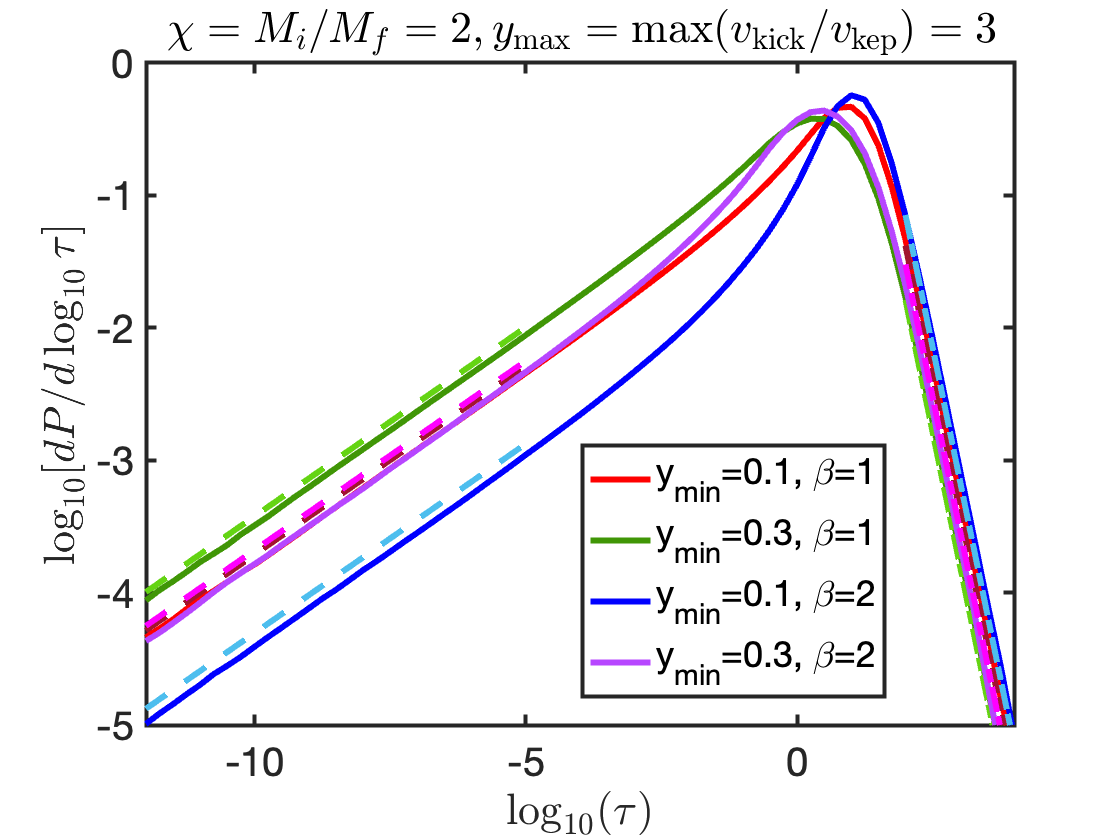}
		\includegraphics[scale=0.2]{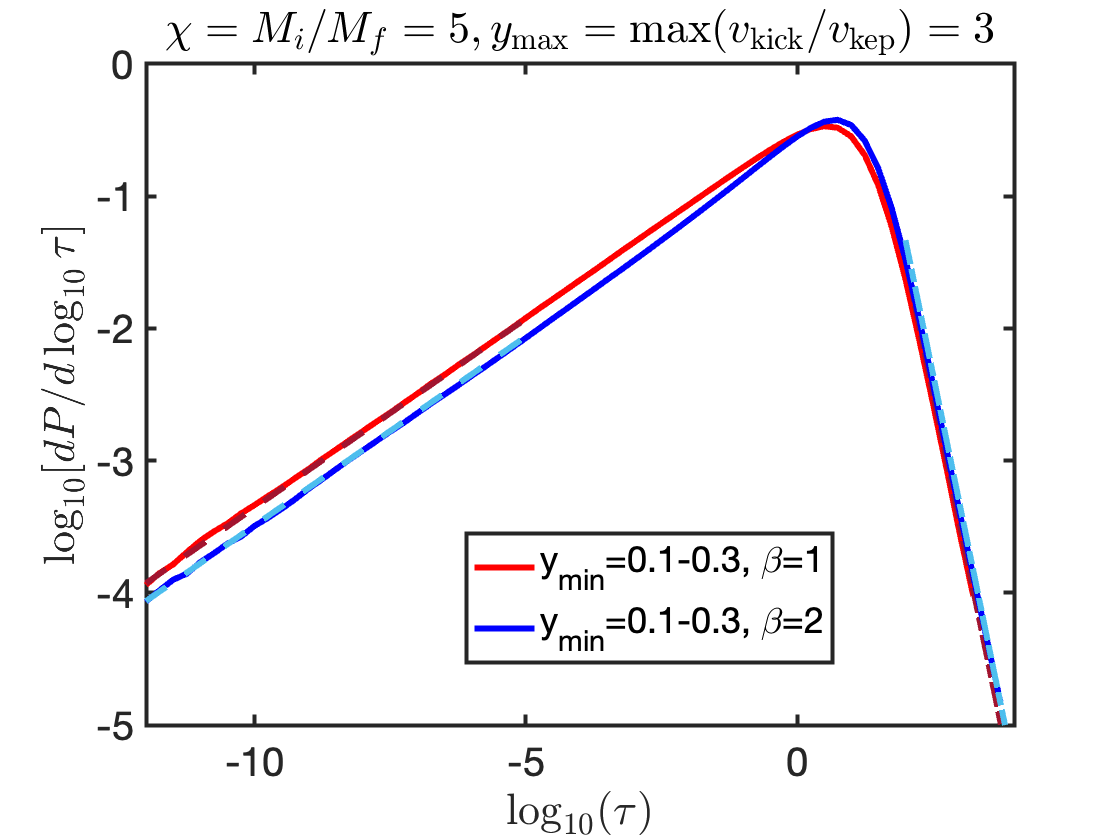}
		\caption{\small The delay time ratio ($\tau$) distributions resulting from kicks with a power law $y=v_{\rm kick}/v_{\rm kep}$ distribution, $dP(y)/dy\propto y^{-\beta}$ for $y> y_{\rm min}$, and random orientations relative to the binary orbit for different values of $y_{\rm min}$, $\beta$ and $\chi=M_i/M_f$. The distributions are normalized relative to the number of bound systems in each case. The left panel depicts results for $\chi=2$ and the right panel depicts results for $\chi=5$ (in the latter case the system is disrupted for $y<0.37$, $y>1.6$ so the distributions are identical for $y_{\rm min}=0.1$ and $y_{\rm min}=0.3$). Dot-dashed lines show the asymptotic scaling for fast mergers (Eq. \ref{eq:PLdistfast}) and dashed lines show the asymptotic scaling for slow mergers (Eq. \ref{eq:PLdistslow}). Specifically, in all cases we recover a $\tau^{2/7} $ distribution at the low end and $\tau^{-2}$ at the high end.}
		\label{fig:comparePLy}
	\end{figure*}

	For the slow merger distribution, the analysis is similar. As for the fast mergers, we integrate over the range of $\delta$ values corresponding to slow mergers:
	\begin{eqnarray}
		\label{eq:PLdistslow}
		& \frac{dP}{d\log \tau}\left(\tau\gg 1\right)\!=\! \left.\frac{dP}{d\delta}\right \vert_{\delta=\chi^{-1}}\int_{\chi^{-1}}^{\sqrt{2/\chi}}\left.\frac{dP}{d\log \tau}\right \vert_{\delta, \tau \gg 1}d\delta\nonumber \\ &\approx \frac{90}{\chi^{3/2}}\left.\frac{dP}{d\delta}\right \vert_{\delta=\chi^{-1}}\tau^{-2}\approx \frac{40}{\chi}P(\frac{1}{\chi}<|\delta|<\sqrt{\frac{2}{\chi}}) .
	\end{eqnarray}
	We see that $dP/d\log \tau \propto \tau^{-2}$ with a normalization (as above) that is proportional to $\chi^{-3/2}$ as anticipated from the overall survival probability (see \S \ref{sec:disrupt}). We note that, due to the different normalizations of Eq. \ref{eq:PLdistfast}, \ref{eq:PLdistslow}, the transition between the two regimes could in principle involve an intermediate region where $dP/d\log \tau$ changes rapidly.
	Fig. \ref{fig:comparePLy} depicts a comparison of the resulting distributions for $dP/dy\propto y^{-\beta}$ with varying values of $y_{\rm min}, \beta, \chi$. The results match well with the asymptotic scaling discussed above.
	
	\section{Post-collapse distributions of semi major axis,  eccentricity and Misalignment}
	\label{sec:aandeandalpha}
	The approach applied in \S \ref{sec:Mergerdelay} to derive the $\tau$ distribution, can be applied to derive the distributions of the semi-major axis, the eccentricity immediately after the collapse and the spin-orbit misalignment.
	The first is strongly related to the time delay distribution. The last two may have a noticeable effect on the GW radiation signal of the merger. 
	
	\subsection{The semi-major axis and eccentricity}
	\label{sec:aande}
	The semi-major axis and the eccentricity  distributions will be gradually modified as the orbit begins to decay due to GW emission. We discuss the initial distribution first and then consider the possible residuals of the eccentricity when the BNS enters the  GW detectors' band.
	
	As discussed in \S \ref{sec:deltadelay} (see in particular the discussion regarding slow mergers), the semi-major axis ratio, $\mathcal{A}\equiv a_f/a_i$ is given by $\mathcal{A}=[2\chi y(\mu_0-\mu)]^{-1}$. Therefore, if $y$ is broadly distributed around $y=1$\footnote{In this context, a broad distribution of $y$ means that $dP/dy$ changes by a factor of $\lesssim 10$ between $y=1$ and $y=1+\min(\sqrt{2}-1,(2/\chi)^{1/2})$. This holds very well for any distribution of $y$ that can be represented as a power-law or as a Gaussian centered at $y\approx 1$ that is not extremely narrow (i.e. $\sigma_{\log_{10}y}\nll 0.3$).}, then
	\begin{equation}
		\label{eq:adist}
		\frac{dP}{d\mathcal{A}}=\frac{dP}{d\mathscr{m}}\frac{d\mathscr{m}}{d{\mathcal{A}}}\approx \frac{1}{\chi^{3/2}} \left.\frac{dP}{dy}\right \vert_{y=1}\mathcal{A}^{-2} \quad \mbox{ for } \mathcal{A}\gtrsim 1.
	\end{equation}

	The eccentricity post-collapse is given by Eq. \ref{eq:esquared}, which, using Eq. \ref{eq:Jratio} gives $1-e^2=(j_f/j_i)^2\chi \mathcal{A}^{-1}$. To examine the distribution of $e$ we can consider two limits. In the case of a slow merger, we have shown that $j_f\sim j_i$ and hence $1-e^2\propto \mathcal{A}^{-1}$. As a result, $dP/de\propto (dP/d\mathcal{A}) (d\mathcal{A}/de)\propto e$. Similarly, in the case of a fast merger, we have shown that $1-e^2\propto (j_f/j_i)^2\propto \tau^{2/7}$. This leads to  $dP/de\propto (dP/d\tau) (d\tau/de)\propto e$. In both cases, the result is the same. Therefore we find that to a first approximation (if $y$ is broadly distributed around $y=1$)
	\begin{equation}
		\label{eq:edist}
		\frac{dP}{de}\approx \frac{6}{\chi^{3/2}} \left.\frac{dP}{dy}\right \vert_{y=1} e \quad \mbox{ for } e\lesssim 0.5\ .
	\end{equation}
	In particular, since for fast mergers $\mathcal{A}\approx 1, 1-e\ll 1$, the peri-center radius $r_p=a_f(1-e)$ becomes proportional to $1-e^2$ and therefore $dP/d\log r_p\propto r_p$.
	
	It is important to stress that $\mathcal{A},e$ are co-dependent. 
	In particular, for a given value of $\mathcal{A}$, there is an upper limit on $1-e^2$. This can be seen by using Eq. \ref{eq:aiaf} to write $\mu(\chi,y,\mathcal{A})$. This is then plugged back into Eq. \ref{eq:esquared} to get $e^2(\chi,y,\mathcal{A},\phi)$. The value of $1-e^2$ can then be shown to be maximal when $\sin^2 \phi=1$ which leads to the limit
	\begin{equation}
		\label{eq:elimit}
		1-e^2<2\mathcal{A}^{-1}-\mathcal{A}^{-2}.
	\end{equation}
	Interestingly, this limit is independent of $y, \chi$.
	Note that this is only the case when $y$ is broadly distributed around $y=1$. If $y<-1+\sqrt{2/\chi}$ then $e$ is always small (see Eq. \ref{eq:esquared} which corresponds to $e\lesssim (\chi-1)+2y$ for small $\chi-1$ and $y$).
	
	Since $\tau \propto \mathcal{A}^4(1-e^2)^{7/2}$ (Eq. \ref{eq:tmerg}), it follows that lines of fixed $\tau$ in the $\mathcal{A}-e$ plane correspond to $\mathcal{A}\propto (1-e^2)^{-7/8}$. At a given time after the collapse, the distribution below this line will have merged. Furthermore, as systems evolve owing to GW losses their eccentricity and semi-major axis both shrink, while following the relation $a\propto e^{12/19}/(1-e^2)$ \citep{Peters1964}.
	
	A limiting condition for GW detectors to detect a binary compact object, before its merger is that its angular frequency is large enough to reach  the detector bandwidth (e.g., roughly 10Hz for LIGO). Since the latter is approximately proportional to $r_p^{-3/2}$, the condition can be translated to an upper limit on the peri-center radius $r_p<r_{\rm p,lim}$ for a binary to be detectable. Considering the distributions of $r_p$ immediately post collapse, there is a small fraction, $N_{\rm GW,0}$, of systems which obtain an extremely high eccentricity (with $1-e\lesssim r_{_{\rm GW}}/a_i$ where $r_{_{\rm GW}}\approx [GM_f/\pi^2 f_{_{\rm GW}}^2]^{1/3}$ is the required peri-center radius for the system to enter the upper frequency end, $f_{_{\rm GW}}$, of the GW detector bandwidth) and satisfy this condition immediately after the collapse. The fraction is then given by
	\begin{equation}
		\label{eq:NGW0}
		N_{\rm GW,0}\approx  \frac{6}{\chi^{3/2}}\left.\frac{dP}{dy}\right \vert_{y=1}\frac{r_{_{\rm GW}}}{a_i}.
	\end{equation}
	
\begin{figure}
		\centering
		\includegraphics[scale=0.2]{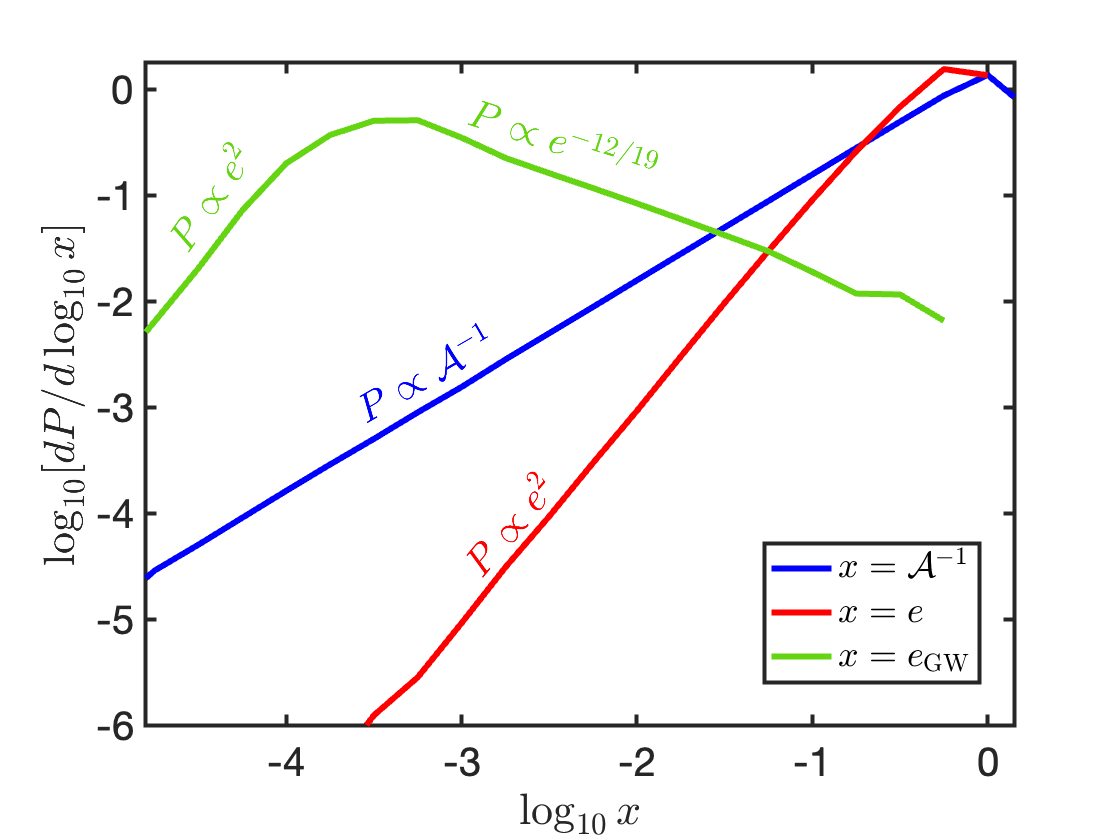}\\
		\includegraphics[scale=0.2]{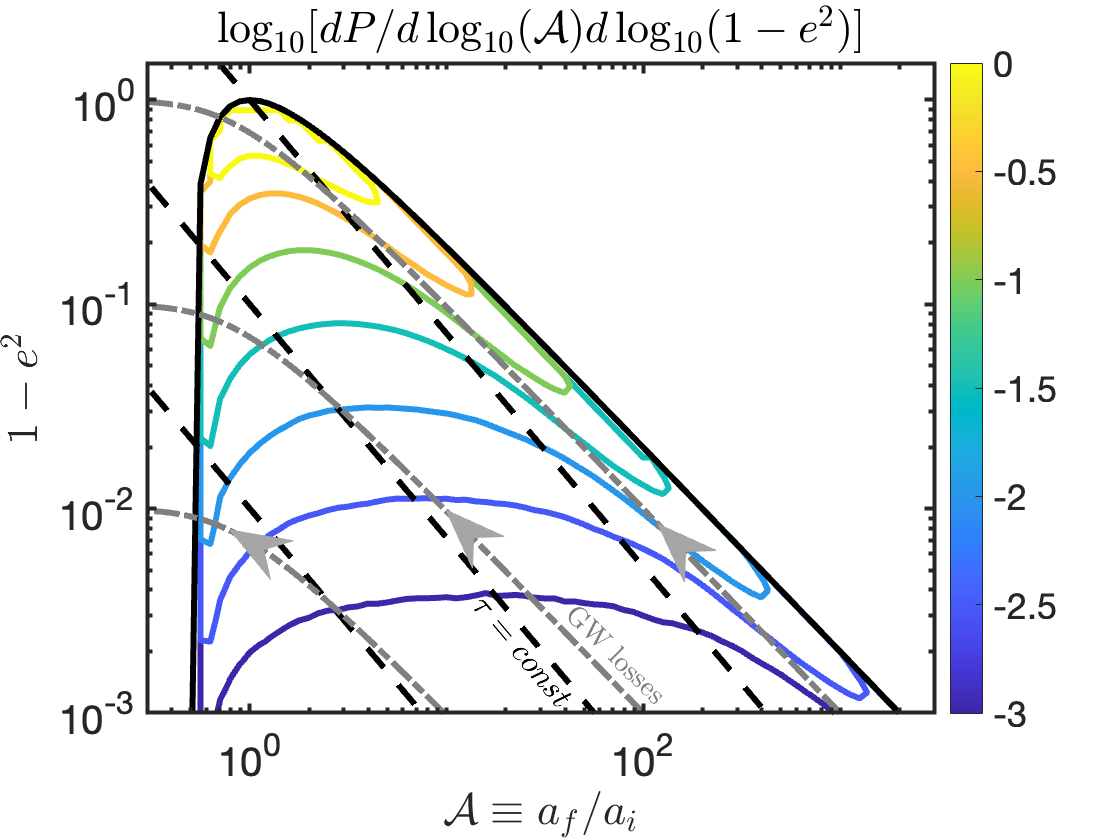}
		\caption{\small Top:  The distributions of $e$ (red; marginalized over $\mathcal{A}$) and $\mathcal{A}^{-1}=a_i/a_f$ (blue; marginalized over $e$). The green line shows the distribution of the eccentricity of the systems as they enter the GW detector range (taking $r_{_{\rm GW}}=0.01a_i$ which is large compared to realistic values, but demonstrates the turnover of the distribution discussed in the text). Results are shown for $\chi=M_i/M_f=2$, and for a power-law distribution in $y=v_{\rm kick}/v_{\rm kep}$ with $y_{\rm min}=0.3, y_{\rm max}=3$ and $\beta=1$. These distributions match the analytic asymptotic scalings given in Eqns. \ref{eq:adist}, \ref{eq:edist}, \ref{eq:NGW0} and \ref{eq:eGW}. Bottom: Contour lines of the  (logarithm of the) 2D probability distribution function $dP/d\log_{10}(\mathcal{A})d\log_{10}(1-e^2)$ describing the systems immediately after the collapse. The resulting values immediately after the collapse are bound by the limit given in Eq. \ref{eq:elimit} and shown by a black solid line. Dashed lines depict constant merger time ratios, $\tau$ (with faster mergers below these lines and slower mergers above). Dot-dashed lines show the evolution of {individual} systems due to GW losses as their orbits become more circular while  gradually shrinking (i.e. moving from bottom right to top left).}
		\label{fig:eanda}
	\end{figure}

	Clearly, the vast majority of systems (which are not disrupted by the collapse) will initially be outside the band of GW detectors. However, as their orbits shrink due to GW radiation, they will eventually become detectable. Since the eccentricity of the systems decays quickly, this population will be characterized by $e_{_{\rm GW}}\ll 1$ as they enter the GW band, which can be approximated by
	\begin{equation}
		\label{eq:eGW}
		e_{_{\rm GW}}\approx \left( \frac{r_{_{\rm GW}}}{a_f}\right)^{19/12} \frac{e}{(1-e^2)^{19/12}} .
	\end{equation}
	Eq. \ref{eq:eGW} provides a good approximation for both $e\ll 1$ and $e\approx 1$ \citep{Peters1964}. It can be used to obtain the eccentricity distribution of GW detectable systems. For systems with $e\lesssim 0.5$ immediately post-collapse, since most systems have $\mathcal{A}\approx 1$, we get $dP/de_{_{\rm GW}}\propto dP/de\propto e_{_{\rm GW}}$. Alternatively, for $e\gtrsim 0.5$ we have $dP/de_{_{\rm GW}}\propto (dP/dr_{p,f})(dr_{p,f}/de_{_{\rm GW}})\propto e_{_{\rm GW}}^{-31/19}$ (see also \citealt{LBB2021}). The transition between these two regimes is at $e\approx 0.5$ which approximately corresponds to $e_{_{\rm GW}}\approx 0.3 (r_{_{\rm GW}}/a_i)^{19/12}$. The $e-\mathcal{A}$ correlation and the probability distributions of $a,e$ are presented in Fig. \ref{fig:eanda} and reproduce well the analytic results described in this section. 
	
	\subsection{Kick misalignment}
\begin{figure}
		\centering
		\includegraphics[scale=0.2]{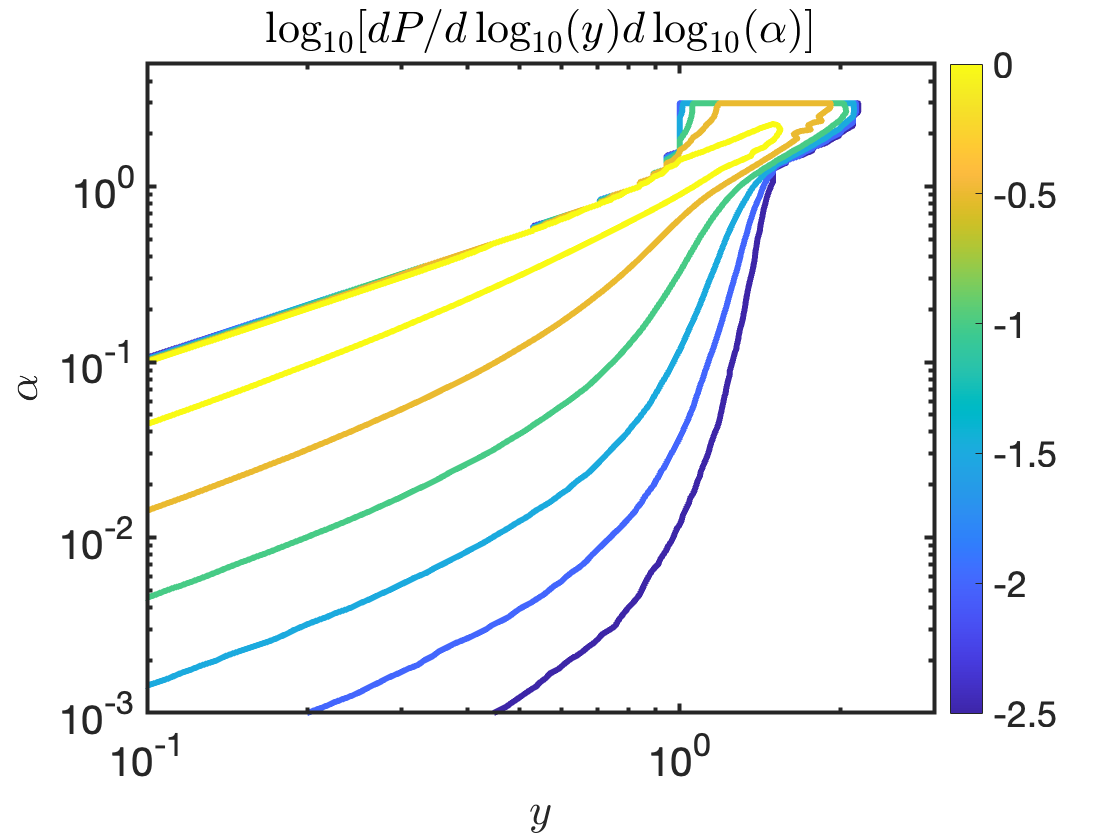}
		\caption{\small Contour lines describing the probability of the degree of misalignment between the final and initial orbital planes, $\alpha$, due to a velocity kick, for $\chi=M_i/M_f=2$ in the ($y=v_{\rm kick}/v_{\rm kep}, \alpha$) plane. The contours depict the (logarithm of the) 2D probability distribution function $dP/d\log_{10}yd\log_{10}\alpha$ describing the systems immediately after the collapse. }
		\label{fig:misalign}
	\end{figure}

	Due to the velocity kick suffered by the collapsing star, the plane of the remaining binary's orbital plane (when it survives) becomes misaligned relative to the initial one. We denote this misalignment angle by $\alpha$. In cases where the spin of either the collapsing star or the companion is aligned with the initial angular momentum, $\alpha$ can be directly measured. $\alpha$ can be calculated as follows (see also \citealt{Hills1983,Kalogera1996,Postnov2014}),
	\begin{equation}
		\label{eq:misalign}
		\cos \alpha= \frac{\vec{J}_{\rm f}\cdot \vec{J}_{\rm i}}{J_{\rm f} J_{\rm i}}=\frac{1+y \mu}{\sqrt{y^2(1-\mu^2)\sin^2\phi +(1+y\mu)^2}}
	\end{equation}
	At $y\ll 1$, this leads to $\alpha\approx \sqrt{2(1-\cos \alpha)}\approx y\sqrt{1-\mu^2}|\sin\phi|$, i.e. the value of $\alpha$ is linear with $y$ (and the mean is $\langle \alpha \rangle \approx y/2$). This persists until $y\sim 1$, at which point if $\mu\approx -1$ then $\alpha$ becomes of order unity. We note also that while Eq. \ref{eq:misalign} does not explicitly depend on $\chi$, it does so in an implicit way, since as $\chi$ increases the range of $\mu$ (and $y$) values that lead to non-disrupted binaries decreases (see Eq. \ref{eq:mu0} and note that $\mu_0$ decreases with $\chi$ for a fixed $y$). The values of $\alpha$ as a function of $y$ and for $\chi=2$ are depicted in Fig. \ref{fig:misalign}.
	
	\section{Typical parameters for pulsars and binary neutron stars}
	\label{sec:paramspace}
	We review here, briefly, several observed properties of BNSs as well as of individual pulsars that are relevant to our analysis. We then discuss their implications for the present study.
	
	\begin{itemize}
		\item {\bf Two BNS populations:} The orbital motion of BNS reveals information on their formation mechanisms and progenitors. The observed population is composed of two types \citep{BP2016,Tauris2017}. In the majority ($\approx 70\%$), the second collapse was bare (with minimal mass ejection, about $0.1 M_\odot$, and a very weak kick). Such a collapse could arise in an electron capture supernovae in which the progenitor has about $1.5\!-\!1.6 M_\odot$. The double pulsar PSR J0737-3039A/B, is a prototype of this population. The rest of the BNS systems form in a regular supernova explosion with a significant mass ejection in a configuration that, without a natal kick, would have become unbound. The first binary pulsar PSR B1913+16 is a prototype of this population. In the following, we will focus on this second population.
		\item {\bf Survival fraction:} There are $\sim 3000$ known pulsars in the Galaxy (as reported in the ATNF catalog; \citealt{ATNF}). 
		The vast majority of pulsars are not in binaries. So far, 19 confirmed BNS have been observed. Assuming that at least half of the pulsar progenitors were in binaries (that have been disrupted)\footnote{As consistent with observations of massive stars \citep{Sana2012}.}, the fraction of surviving binaries is about 1-2\%. The fraction of surviving binaries, that formed within regular supernovae is smaller. The exact value, which may be closer 0.5\% is not important for the subsequent discussion. What matters is that this overall survival fraction is very small, regardless of whether this is dominated by the first or second collapse.
		
		
		\item {\bf Pulsar velocities:} The young pulsar mean velocity (as estimated from their projected proper motions) is $\approx 400-500\mbox{km s}^{-1}$ \citep{Hobbs2005,Kapil2023}. These estimates are based on $\sim 80$ pulsars with young characteristic ages and therefore it is quite robust. These authors also found that few pulsars even have (2D projected) velocities as large as $1600\mbox{km s}^{-1}$.
		\item {\bf CM motion:} {The center of mass (CM) velocities of BNS systems with respect to the local standard of rest can be estimated from their proper motion and inferred distances. Such estimates are available for only three of the BNS systems which have a high probability of having formed via a ``regular" SN: B913+16, B1534+12 and J1757-1854 \citep{Weisberg2005,Fonseca2014,Cameron2023}. The average velocity estimated for those systems is $150-200\mbox{km s}^{-1}$.} 
	\end{itemize}

	A very conservative upper limit on $\chi$ can be obtained from the survival probability. As explained above, we expect that the survival fraction during the second collapse (bound from below by the overall survival fraction), $P_{\rm sur}$ for systems formed within the regular supernova channel is $P_{\rm sur}\gtrsim 0.01$. Since for $\chi\gg 1$, there is only a narrow range of width $\Delta y\approx 2\sqrt{2/\chi}$ within which systems might survive a supernova (with a probability of $\lesssim \chi^{-1}$, see Eq. \ref{eq:Psur} and Fig. \ref{fig:chiyparamspace}). This means that to a good approximation, one has $P_{\rm sur}<\left.\frac{dP}{dy}\right \vert_{y=1}\chi^{-3/2}$. Since the range in $y$ covers at least a factor of a few (see discussion in the next paragraph), we can safely take $\left.\frac{dP}{dy}\right \vert_{y=1}\lesssim 1/3$ which in turn leads to $\chi<10$ to guarantee that $P_{\rm sur}$ is not too small. We stress that this is a limit on the typical value of $\chi$, and this argument does not preclude that rare individual objects might have much larger values. This argument is independent of our knowledge of stellar evolution, and indeed one can improve the limit by taking that into consideration as discussed next.
	Since most stars with initial masses $\gtrsim 8M_{\odot}$, result in NS formation, and since the initial mass function declines rapidly with mass at $M>8M_{\odot}$, a safe upper limit on $\chi$ for a typical BNS before the formation of the 2nd NS is $\chi\lesssim 5$ (corresponding to an exploding star mass of $M\lesssim 12M_{\odot}$). $\chi$ could be significantly smaller if the star is initially smaller or if it losses a significant fraction of its  mass due to winds and binary interaction, prior to its eventual collapse (indeed for bare collapses $\chi-1\approx 0.05$).
	
	We turn now to the distribution of $y$, and argue that it is likely to be wide and, in particular, $y\approx 1$ should be relatively common. From observed BNS we can relate the the semi-major axis and eccentricity post-collapse (but before significant GW decay) to the initial semi-major axis $a_f(1-e)<a_i<a_f(1+e)$. From this one infers that $a_i$ spans a range at least as wide as $10^{11}\mbox{ cm}\lesssim a_i\lesssim 10^{13}$\,cm \citep{BP2016}. For $\chi=3$, this corresponds to $10^2\mbox{km s}^{-1}\!\lesssim \!v_{\rm kep}\!\lesssim \!10^3\mbox{km s}^{-1}$. This range overlaps the range of typical kick velocities as inferred from velocities of young pulsars, $\sim 450$ km s$^{-1}$, discussed above. This suggests that $y\approx 1$ should be common (see Eq. \ref{eq:v2f}). In particular if $y$ were typically $\ll 1$ (i.e. pulsar velocities dominated by the Blaauw kick components, \citealt{Blaauw1961}), then Eq. \ref{eq:v2f} shows that for $\chi\approx 3$, we would have $v_{2f}\approx 0.17v_{\rm kep}$ which would both be unable to account for higher velocity pulsars and would significantly overproduce low velocity pulsars as compared with observations. 
	
	Additional support to this conclusion arises from the center of mass velocity of BNS systems formed by ``regular" SN which is not much smaller than the Keplerian velocity of those systems.  Using Eq. \ref{eq:delvcm} we see that this rules out the possibility that typically $y<-1+\sqrt{2/\chi}$  for the bound systems. Furthermore, it would not be possible to account for this sub-population of BNS (which are typically characterized by large $e$, see \citealt{BP2016}) if $y\ll 1$ during the collapse leading to a BNS (see Fig. \ref{fig:chiyparamspace}).
	
	We conclude that during BNS formation, $1\leq \chi \lesssim 5$ and $y$ spans a wide distribution of values around $y=1$. Such a distribution is required to account for the basic observations of pulsars and BNS systems \footnote{Note we do not attempt here to estimate the exact distributions of $y$ and $\chi$ and any correlations that may exist between the various parameters. Such an attempt would clearly have to tackle the complex interactions and evolutionary paths involved. Indeed, it is exactly because of the richness of the underlying stellar evolution that we are focused here on deriving general but also robust constraints on the $y,\chi$ parameter space.}. This means that high eccentricity and therefore rapidly merging systems are a non negligible outcome of the collapse. In \S \ref{sec:bns} we discuss the implications of our analysis described in \S \ref{sec:Mergerdelay} to the population of BNSs.
	
	\section{Application to binary neutron star mergers}
	\label{sec:bns}
	
\begin{figure}
		\centering
		\includegraphics[scale=0.2]{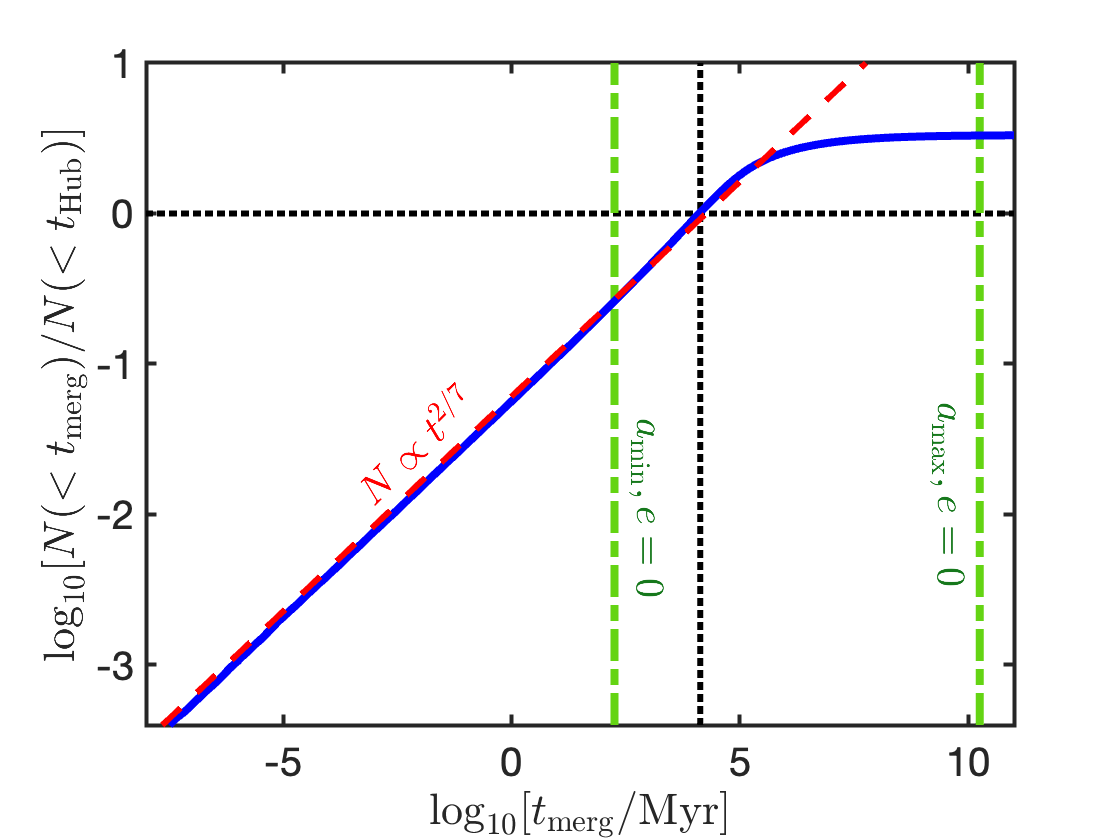}
		\includegraphics[scale=0.2]{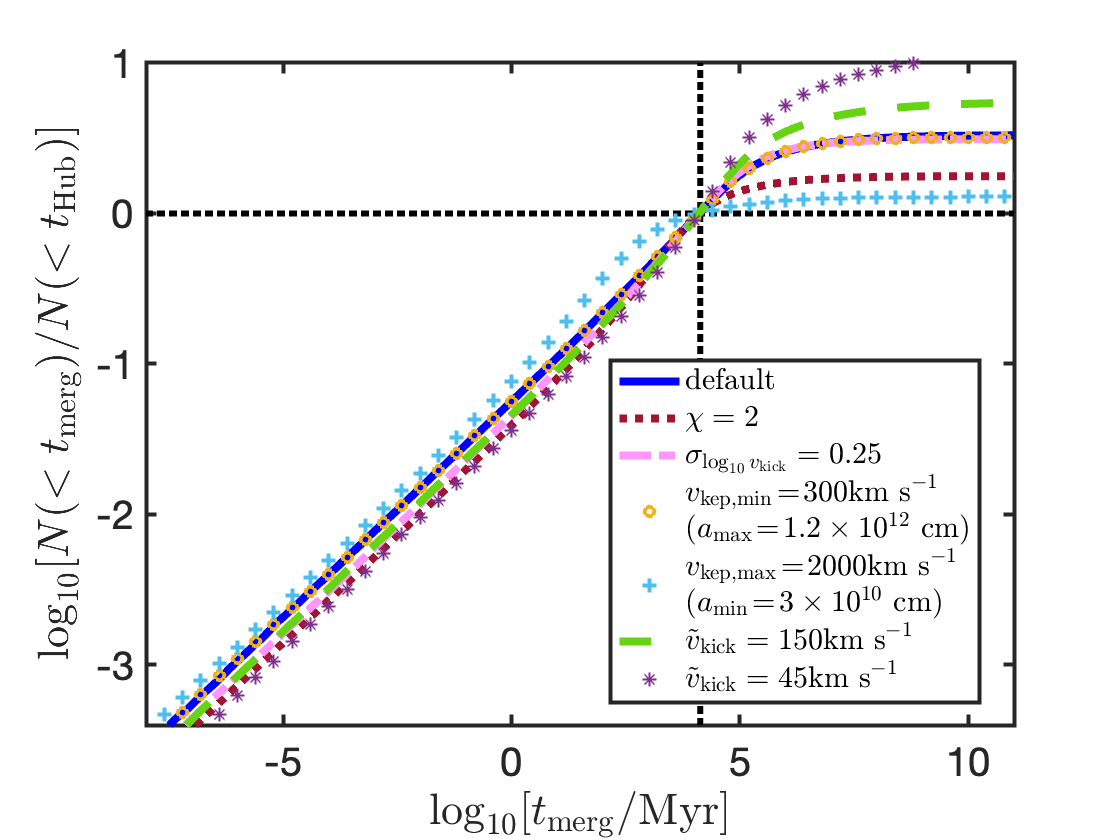}
		\caption{\small  Top: Fraction of bound systems with a merger time less than $t_{\rm merg}$ (normalized relative to the fraction of systems merging within a Hubble time, as visualised by the dotted lines), for a population of BNS progenitors with zero initial eccentricity and  $dN/dv_{\rm kep}\propto v_{\rm kep}^{-1}$ between $v_{\rm kep,min}=100\mbox{km s}^{-1}$ and $v_{\rm kep,max}=1000\mbox{km s}^{-1}$ (for the assumed initial mass this is equivalent to $dN/da_i\propto a_i^{-1}$ between $a_{\rm min}=1.2\times 10^{11}\mbox{ cm}$ and $a_{\rm max}=1.2\times 10^{13}\mbox{ cm}$). The fractional mass loss during the collapse is $\chi=M_i/M_f=3$ and the kick velocity is log-normal distributed with a median value of $450\mbox{km s}^{-1}$ and $\sigma_{\log_{10}v_{\rm kick}}=0.5$.  A solid line shows the results from a Monte Carlo simulation and a dashed line shows the asymptotic power-law tail, $N=0.06(t/\mbox{Myr})^{2/7}$ as given in Eq. \ref{eq:tmergcum}. Dot-dashed green lines show the merger times for BNSs with the same total masses and with circular orbits with separations bound between the pre-collapse values (between $a_{\rm min}$ and $a_{\rm max}$). Bottom: The same distribution, but with each of the following changes relative to the default case shown above: (i) lower initial total mass, (ii) narrower kick velocity distribution, (iii) lower maximal initial separation and (iv) lower initial minimal separation and (v) lower median kick velocity (two different cases shown). All cases show the $t^{2/7}$ tail at short merger times and an almost identical  fraction of systems with $t_{\rm merg}<1$\,Myr.}
		\label{fig:tmerdist}
	\end{figure}

	Consider, now,  a typical BNS progenitor system with, $M_i=9M_{\odot}$, a final total mass of $M_f=3M_{\odot}$ and $\chi=3$. Such a system is a prototype of a BNS of the type that we describe and the characteristic $\chi$ doesn't change significantly when considering other relevant systems (excluding bare collapses).
	In addition, we assume the distribution of Keplerian velocities  to be $dN/dv_{\rm kep}\propto v_{\rm kep}^{-1}$ with $v_{\rm kep,min}=10^2\mbox{km s}^{-1}$ and $v_{\rm kep,max}=10^3\mbox{km s}^{-1}$. For the initial mass assumed above, this corresponds to semi-major axes of the pre-collapse binary distributed as $dN/da_i\propto a_i^{-1}$ between $a_{\rm min}=1.2\times 10^{11}\mbox{cm}$ and $a_{\rm max}=1.2\times 10^{13}\mbox{ cm}$.
	Finally, motivated by the velocity distribution of young pulsars \citep{Hobbs2005}, we take the kick velocity to be a log-normal distribution with a median value of $450\mbox{km s}^{-1}$ and $\sigma_{\log_{10}v_{\rm kick}}=0.5$ (see also typical velocities found by \citealt{Kapil2023}, although note that in their modelling the velocity depends on the pre-collapse properties). The resulting merger time distribution is shown in Fig. \ref{fig:tmerdist}. At low $t_{\rm merg}$, the asymptotic distribution is proportional to $t_{\rm merg}^{2/7}$ as shown in \S \ref{sec:ydist}. In particular,
	\begin{equation}
		\label{eq:tmergcum}
		\frac{N(t_{\rm merg}'<t_{\rm merg})}{N(t_{\rm merg}'<t_{\rm Hub})}=0.06\left(\frac{t_{\rm merg}}{\mbox{Myr}}\right)^{2/7} \quad \mbox{ for } t_{\rm merg}\lesssim 100\mbox{ Gyr.}
	\end{equation}
	At larger $t_{\rm merg}$, the probability is dominated by the initial separation distribution. At this limit, the merger distribution increases logarithmically, since $dN/da_i\propto a_i^{-1}$ (and $e=0$) leads to $dN/dt_{\rm merg,i}\propto t_{\rm merg,i}^{-1}$,  which is the commonly used $t^{-1}$ time delay distribution \citep{Piran1992}. Importantly, as explained in \S \ref{sec:ydist}, the delay time distribution at short merger times, is extremely robust to the specifics of the assumed underlying distributions. To demonstrate this, we show, in the bottom panel of Fig. \ref{fig:tmerdist}, the results obtained for a similar analysis but with each of the following changes: (i) lower initial total mass ($\chi=2$), (ii) narrower kick velocity distribution ($\sigma_{\log_{10}v_{\rm kick}}=0.25$), (iii) lower maximal initial separation ($a_{\rm max}=1.2\times 10^{12}\mbox{ cm}$) and (iv) lower initial minimal separation ($a_{\rm min}=3\times 10^{10}\mbox{ cm}$) and (v) lower median kick velocity, either $\tilde{v}_{\rm kick}=150\mbox{km s}^{-1}$ or $\tilde{v}_{\rm kick}=45\mbox{km s}^{-1}$. All cases show the $t^{2/7}$ tail at short merger times and an almost identical fraction of systems with $t_{\rm merg}<t_{\rm Hub}$. There are two minor exceptions to this. First, case (iv), for which, even without kicks, the merger time can be as short as $\sim1$Myr, and consequently the fraction of systems with $t<t_{\rm Hub}$ is even greater than the other cases. Second, case (v) with $\tilde{v}_{\rm kick}=45\mbox{km s}^{-1}$ (lower by an order of magnitude from the default case). In this case the fraction of systems with ${v}_{\rm kick} \approx v_{\rm kep}$ (or equivalently $\delta\lesssim \chi^{-1}$, see Eq. \ref{eq:PLdistfast}) is reduced, and as a result the fraction of systems with $t<t_{\rm Hub}$ is slightly smaller (by a factor $\lesssim 2$) than the other cases.

	The overall probability of a BNS system to merge within a Hubble time, for these distribution parameters, is $N(t_{\rm merg}'<t_{\rm Hub})\!\sim\! 1/3$. Among the merging systems the time delay distribution is $dN/dt_{\rm merg}\propto t_{\rm merg}^{-5/7}$. This distribution is harder than the ``classical" $t^{-1}$, but it extends down to much shorter timescales which without kicks would require unreasonably small initial separations or extremely large eccentricities. Recall that for $a=10^{11}$cm, which is a reasonable lower cutoff for the initial separation, the merger time is: $\sim 100$\,Myr. On the other hand, Eq. \ref{eq:tmergcum} implies that $\sim 11\%$ of the merging BNSs do so within 10\,Myr and $\sim 1.6\%$ do so within 10kyr.

	We see that, generically, fast mergers must be a non-negligible fraction of ``regular"-collapse BNS systems. This result is robust, and it arises from very conservative assumptions on the properties of the immediate progenitors of BNS systems (see \S \ref{sec:paramspace}), combined with the generic shallow asymptotic behavior, $\propto t^{2/7}$, of the fraction of systems merging within $t$.

	In \cite{BP2019} we have shown that the observed sample of Galactic BNSs requires that at least $\sim 40\%$ of BNS systems merge within less than a Gyr. Furthermore, we have shown that this population is dominated by systems formed with low eccentricity and is therefore a reflection of the small separation of a significant sub-population of BNS progenitors just before the collapse and not of kicks (see also \citealt{Belczynski2006}). This population is most likely that of ``bare" collapses, which, as discussed in \S \ref{sec:paramspace}, constitute the majority of BNS systems.
	At the same time, while typically faster to merge, the merger time due to bare collapses is limited by the smallest possible initial separation. Recalling that for $y\ll1 , \chi-1\ll 1$ we have $t_{\rm merg}\approx 0.1 a_{i,11}^4$\,Gyr, we see that $t_{\rm merg}<1$\,Myr requires $a_{i}<3\times 10^{10}$\,cm or less than half of a solar radius. This is substantially lower than expected to be generically possible owing to common envelope evolution \citep{Kruckow2016}. Therefore, at sufficiently low merger times, the distribution should become dominated by fast mergers due to kicks in ``regular" collapses as discussed in this work. 
	
	Finally, we combine our results regarding the short delay tail of the merger time distribution obtained from ``regular" collapses with our results from \cite{BP2019} for the delay time distribution, which were dominated by the ``bare" collapse channel. The overall (regular+bare collapse) BNS delay time distribution (including systems merging on $t>t_{\rm Hub}$) can be roughly approximated as
	\begin{eqnarray}
		\frac{dN}{d\log t}\!\approx\! \left\{\begin{array}{ll}5\times 10^{-3}\left(t/30{\rm Myr}\right)^{2/7} & t\lesssim 30{\rm Myr}\,\\
			0.2\left(t/30{\rm Myr}\right)^{-0.3} & 30\mbox{Myr}\!\lesssim\! t \!\lesssim\! 15\mbox{Gyr}\,\\
			0.031 & \mbox{ else}\ .
		\end{array} \right.
	\end{eqnarray}
	We note that, due to the presence of the two populations, there is a very rapid transition at $t\approx 30$\,Myr from bare collapses with small separation (above this delay time) to ultrafast mergers from the regular collapse channel (below this delay time). To a good approximation, the overall distribution can be modeled as discontinuous at 30Myr.

	\subsection{Implications for GW detectors}
	\label{sec:GWdetectors}
	A BNS enters the frequency range detectable by LIGO when its peri-center radius becomes $r_{_{\rm GW}}\approx 7\times 10^7 (M_f/3M_{\odot})^{1/3}$\,cm. The cumulative fraction of bound systems (from the ``regular" collapse channel) with a given $e_{_{\rm GW}}$ as they enter the GW detectors'  band is
	\begin{eqnarray}
		N(>e_{_{\rm GW}})\approx 10^{-4}\times 
		\!\left\{ \begin{array}{ll}\left(\frac{e_{_{\rm GW}}}{0.3}\right)^{-12/19} & 1.4\times10^{-7}\lesssim e_{_{\rm GW}}\lesssim 0.3\,\\
			1 & \mbox{ else}\ .
		\end{array} \right.
	\end{eqnarray}
	in accordance with Eq. \ref{eq:NGW0} and the power law scaling for lower $e_{_{\rm GW}}$ discussed in \S \ref{sec:aande}. These events will have a distinctive GW chirping signal that will distinguish them from the more common circular mergers.
	
	A smaller fraction of events would have a finite eccentricity even up to the point of contact between the NSs. Taking the separation at the point of contact between the NSs to be $\sim 2\times 10^6$\,cm, there is a fraction $N(>e)\approx 3\times 10^{-6} (e/0.3)^{-12/19}$ of systems with eccentricities greater than $e$ at that point. Eccentric mergers of this kind likely affect the properties of the mass ejection and the size of the debris disk formed as a result of the merger. Although the quantitative effect of this is still under debate, general relativistic numerical simulations of such mergers have generally reported on a tendency for both the ejected mass and the disk mass to be larger for eccentric mergers as compared with circular ones \citep{Gold2012,East2012,Radice2016,East2016, Chaurasia2018}.

	Space-based GW detectors will be sensitive to much lower frequencies than LIGO. For instance, the approximate frequency range of LISA is expected to be $0.1-100$\,mHz.
	In that case, we find that systems enter the GW band with a peri-center of $r_{_{\rm GW}}\approx 3\times 10^{10} (M_f/3M_{\odot})^{1/3}(f_{_{\rm GW}}/\mbox{mHz})^{-2/3}$\,cm. In other words, $r_{_{\rm GW}}$ is comparable to $a_{\rm min}$, and a large (i.e., tens of percent) fraction of BNS systems formed by ``regular collapse" should have a significant and detectable eccentricity. Such detectors would be able to probe the eccentricity distribution immediately after the collapse ($N(<e)\propto e^2$) and, importantly, would be able to clearly separate between the regular and bare collapse channels (recall that for the latter, the eccentricity is limited to $e\lesssim (\chi-1)+2y\ll 1$). In addition, \cite{Lau2020}, have pointed out that most LISA-detected BNSs would be localized to $\lesssim 2^{\circ}$, and as a result, for Galactic objects, their height above the Galactic plane can be measured and limits on their kick velocity received at birth could be obtained.
	We stress that the idea of learning about BNS kicks from space-based GW interferometers is not new. Previous works have explored, using a population synthesis approach, the number and eccentricities of BNS in-spirals that would be detectable by LISA \citep{Nelemans2001,Belczynski2010,Vigna-Gomez2018,Kyutoku2019,Lau2020,Andrews2020}. The specific predictions, depend on the assumed birth properties of BNS binaries and can differ by a factor of up to ten. The analytic approach presented here enables us to easily explore different birth property statistics.
	
	An additional signature of the kick induced rapid merger delay tail, is that it correlates with a large degree of misalignment between the spin of the newly born NS and the binary orbit. This too (along with the eccentricity), could lead to an imprint on the GW waveform. A difficulty with detecting this imprint is that the magnitude of the spin diminishes significantly with time, as the newly born NS spins down via EM dipole radiation. However, if the mergers are sufficiently rapid and the magnetic field of the freshly born NS is modest, the spin may remain detectable at the time of merger.
	
	\subsection{Implications for r-process enrichment}
	
	If BNS mergers dominate $r$-process production in the Galactic history, then the global late decline of Galactic $r$-process (relative to iron) abundance requires BNS merger delay times of $t_{\rm merg}\lesssim 250$\,Myr to be common \citep{Mennekens2014,HBP2018,Cote2019ApJ,Simonetti2019}. As explained above, that sub-population of BNSs is dominated by the systems with rapid mergers due to small initial separations and is not driven by the population of rapid mergers due to kicks discussed here. {Shorter delay times between star formation and $r$-process enrichment (e.g. associated with shorter GW merger delays as discussed here) could lead to a fraction of Galactic metal poor stars with large Europium abundances \citep{Argast2004,Matteucci2014,ishimaru2015ApJ,BDS2018,Tarumi2021} and to $r$-process enrichment of environments in which star formation is relatively short lived such as globular clusters and ultra faint dwarf stars \citep{Tsujimoto2014,ji2016Nature,BHP2016b,Safarzadeh2019,kirby2020}.}

	Finally, we note that a small fraction of rapid mergers will have a non-negligible eccentricity during the last stages of the merger process (see \S \ref{sec:GWdetectors}). Such mergers could lead to a greater amount of neutron rich ejecta, and therefore to an increased $r$-process yield (with a potentially distinct abundance pattern from circular mergers). Furthermore, rapid BNS mergers will preferentially probe regions with higher external interstellar gas density.
	As a result, the associated Kilonovae and their afterglows from these events will be brighter and more readily detectable, despite the relative rarity of such mergers.
	
	\subsection{Implications for short GRBs}
	
	An implication of the $t^{2/7}$ merger delay tail, is that a fraction of mergers will happen while there is still a low density ``bubble" in the external medium due to the passage of the SN blast-wave from the second SN in the binary. The result will be a short GRB within a SN remnant (SNR). A somewhat analogous situation was discussed by \cite{Margalit2020}, who consider the propagation of the kilonova ejecta within the region evacuated by the GRB jet's propagation.
	While the SNR is unlikely to be directly detectable, it will affect the properties of the observed short GRB, as we briefly summarize next. First, notice that the Thomson optical depth associated with the SN ejecta material is $\tau_{_{\rm T}}\approx 4\times 10^{-5} (M_{\rm ej}/6M_{\odot})(R_{_{\rm SNR}}/\mbox{ pc})^{-2}$ where $R_{_{\rm SNR}}$ is the SNR radius  at the time of the merger. $R_{_{\rm SNR}}$  will typically be of order a few times the Sedov length, which corresponds to a several pc. Since $\tau_{_{\rm T}}\ll 1$, unless the GRB-SN delay is less than a few years, the SN ejecta shell does not affect the propagation of the GRB $\gamma$-rays, and the prompt phase remains unaffected. Rapid mergers may have a different impact on the prompt emission properties. In particular, a fraction of the most rapid mergers will still be eccentric during the merger itself (see \S \ref{sec:GWdetectors}). This may lead to an increased mass in the debris disk formed during the merger, which in turn could extend the duration of the associated prompt emission.

	The afterglow will be strongly affected by the SNR. While the GRB blast-wave is propagating through the region evacuated by the SN, the deceleration of the former will be strongly suppressed. As a result, the afterglow onset will be delayed by $t_{\rm af}\approx R_{_{\rm SNR}}/2c\Gamma^2\approx 5000 (R_{_{\rm SNR}}/\mbox{ pc})(\Gamma/100)^{-2}$\,s (where $\Gamma$ is the initial Lorentz factor of the GRB blast-wave). Furthermore, if the delay between the SN and the merger is comparable or shorter than the Sedov time, the density encountered by the GRB blast-wave once it reaches the SNR is going to be enhanced relative to the ambient external density, leading to more intense afterglow emission once it begins (relative to an afterglow in an unperturbed medium at the same observation time). Furthermore, the short delay time associated with the merger would imply that the BNS could not have travelled far from its birth location by the time it merges, and it would therefore encounter an external density that is larger on average than would be experienced by systems with longer delay times \citep{Duque2020}.

	Overall, a GRB of this type should be identifiable as (i) a short GRB in a star-forming region (although, as mentioned above, the prompt duration may be somewhat extended for particularly eccentric mergers) that (ii) has an afterglow onset (relative to the prompt) delayed by $\sim 1$\,hr and (iii) is typically brighter once it begins (relative to other short GRBs at a similar observation time). 
	
	\subsection{Radio precursors of BNS mergers}
	Based on the Galactic magnetar formation rate, it is estimated that a large fraction ($\sim 40\%$) of NSs start their lives as magnetars \citep{Beniamini2019}. This suggests that one (or both) of the NSs in the binary could initially be a magnetar. The lifetimes of magnetars (before their field decays substantially) are relatively short and are typically estimated to be $10^3-10^4$\,yrs \citep{Dall'Osso2012,Vigano2013}. Therefore, considering a bare collapse channel, the magnetic field would have completely decayed by the time of the merger. However, the ultrafast kick induced delay time tail for regular collapses discussed here, means that there would be a non-negligible fraction of mergers in which one of the NS is strongly magnetized during the merger. Furthermore, as recently shown by \cite{Beniamini2023}, there is growing evidence of a large population of ultra-long period magnetars, whose fields decay on a much longer $\sim 10^6$\,yr timescale, making them favourable candidates for being involved in mergers involving a highly magnetized NS. Several authors \citep{Lipunov1996,Lyutikov2019,Cooper2022} have suggested that such mergers could result in detectable coherent radio precursors of BNS mergers. Detection of such a radio precursor could be of huge importance, as it could enable a distance measurement based on the dispersion measure (and independent of redshift) - this would make even a few such events extremely useful as cosmological probes.

	\section{Summary}
	\label{sec:summary}
	We have studied in this paper the general outcome of a binary 
	composed of a stellar mass (possibly compact) object and a star that undergoes a gravitational collapse. We assume that, due to tidal forces, the binary is initially in a circular orbit. The {main} parameters describing the event are as follows: the initial/final total mass ratio, $ \chi \equiv M_i /M_f $, the kick velocity  given to the remnant relative to the initial Keplerian velocity of the binary, $y\equiv v_{\rm kick}/v_{\rm kep}$, and the angle between the kick velocity and the orbital velocity of the collapsing star ($\theta$ or equivalently $\mu=\cos \theta$).
	We have considered the survival probability of the binary, the resulting GW merger delay time, spin-orbit misalignment, and systemic velocities of either the binary or the escaping compact object due to the collapse.
	Our main findings are as follows:
	\begin{itemize}
		\item When the kick velocity is similar and opposite in direction to the Keplerian velocity ($v_{\rm kick}\approx v_{\rm kep}, \mu\to -1$; and for any $M_i /M_f$), the system  becomes highly eccentric, leading to an ultrafast merger (i.e., with GW merger delay time lower by orders of magnitude compared to that associated with the pre-collapse orbit). These systems also result in an order unity misalignment between the spin of the newly formed compact object and the binary orbit.
		\item For low mass ejection, $M_i<2M_f$, and moderate kick velocity, $v_{\rm kick}/v_{\rm kep}<-1+\sqrt{2M_f/M_i}$, binaries always survive the collapse. In this regime, the typical eccentricity is $e\lesssim \Delta M/M_f+2v_{\rm kick}/v_{\rm kep}$ (where $\Delta M=M_i-M_f$ is the ejecta mass), the misalignment angle is $\alpha \approx v_{\rm kick}/(2v_{\rm kep})$ and the change in the binary's CM velocity is $\Delta v_{\rm CM}\leq v_{\rm kep}(\Delta M/M_f)(M_1/M_i)+v_{\rm kick} (M_{2f}/M_f)$ (where $M_1$ is the mass of the non-collapsing star and $M_{2f}$ is the final mass of the collapsed star), all becoming small, regardless of kick orientation.
		\item For significant mass ejection, $M_i>2M_f$, binaries are disrupted regardless of the kick orientation, either when the kicks are too weak ($v_{\rm kick}/v_{\rm kep} <1-\sqrt{2M_f/M_i}$) or when they are too strong ($v_{\rm kick}/v_{\rm kep} >1+\sqrt{2M_f/M_i}$). The velocity of the newly formed compact object is $\approx v_{\rm kep}(M_1/M_i)+v_{\rm kick}$.
		\item If the kick orientations are isotropically distributed (in the frame of the collapsing star) and $v_{\rm kick}\approx v_{\rm kep}$, the probability of binary survival is $P_{\rm sur}\approx M_f/(2M_{i})$. If $v_{\rm kick}/ v_{\rm kep}$ is not extremely narrowly distributed around $v_{\rm kick}/ v_{\rm kep}=1$ (i.e. the probability associated with the range of $v_{\rm kick}/ v_{\rm kep}$ spans order unity or more), then $P_{\rm sur}\propto (M_i /M_f)^{-3/2}$.
		\item If $v_{\rm kick}/v_{\rm kep}$ is broadly distributed around $1$ (and the kicks are isotropic), then the GW merger delay time distribution inevitably develops a shallow short merger delay time tail: $dP/d\log t\propto t^{2/7}$. The result is a robust population of `ultrafast' mergers extending to arbitrarily short merger times. At long merger times the distribution drops steeply as $dP/d\log t\propto t^{-2}$. The only way to circumvent the existence of such an ultrafast merger population is to have practically no overlap between the kick and Keplerian velocity distributions, which, as discussed above could also result in very high disruption probability, unless the mass loss is modest. 
		\item If the delay time distribution corresponding to no kicks (i.e. arising simply as a result of the spread in the initial orbital parameters of the binaries) is characterized by $dP/d\log t\propto t^{s}$ then the overall delay time distribution (i.e. including kicks) can be roughly approximated by a convolution of the kick ansatz described above with the distribution corresponding to no kicks. This leads to a broken power law form in $d\log P/d\log t$ where the power-law index varies from $2/7$ at low $t$ to $-2$ at large values of $t$. If $-2<s<2/7$, then there will be an intermediate range of $t$ values with a power law index of $s$. For example, for $s\approx 0$ (often invoked for BNS mergers), the fast merger tail ($dP/d\log t\propto t^{2/7}$) will dominate at small values of $t$ (below the minimum delay corresponding to no kick). The distribution will transition to the `no kick' distribution, $dP/d\log t\propto t^s$ at intermediate values and finally to $dP/d\log t\propto t^{-2}$ at merger times beyond the maximum corresponding to no kicks.
	\end{itemize}
	
	We have applied these general results to the formation of BNS systems in which the collapse involved a regular SN (i.e. ignoring the population of BNS formed by ``bare collapses" such as via electron capture SNe in which there is almost no mass ejection and correspondingly a very small kick).  Our main findings are as follows: 
	\begin{itemize} 
		\item The fraction of BNS systems relative to the overall number of pulsars, along with the measured systemic velocities of these respective systems, suggests that before the second collapse BNSs typically have  $1\lesssim M_i /M_f \lesssim 5$ and with a wide distribution of $v_{\rm kick}$ around $v_{\rm kep}$. The implication is that ultrafast mergers are an inevitable outcome of this BNS formation channel. About a percent of all BNS systems should have delay times $t\lesssim 30$\,Myr. The fraction decreases only moderately (as $t^{2/7}$) at shorter times. In particular, kicks lead to a sub-population of BNSs with merger delays that are orders of magnitude shorter than those that can ever be achieved by simply having a small separation before the collapse (e.g. due to a common envelope). Finally, we have demonstrated that this result is largely insensitive to plausible changes in the parameters of the underlying distributions (i.e amount of mass loss, varying initial separation distributions, and varying typical and width of the kick distribution).
		\item Ultrafast BNS mergers can enter a GW detector's frequency range with non-negligible eccentricity. If their mergers are fast enough, their spin-orbit misalignment might also remain measurable. Low frequency GW detectors would be able to distinguish between different channels of collapse (i.e., involving small or large kicks and mass ejection).
		\item Ultrafast mergers could lead to extreme $r$-process enrichment in environments that are metal poor and/or in which star formation is short lived. Both the kilonovae associated with ultrafast mergers and their radio afterglows will be brighter than for regular BNS mergers.
		\item Short GRBs associated with ultrafast mergers can take place while  they are still confined within the low density bubble carved by the SN blast-wave. A GRB of this type should be identifiable as a short GRB in a star-forming region, with a bright afterglow, the onset of which is delayed by $\sim 1$\,hr compared to the prompt.
		\item Ultrafast BNS mergers can take place when the newly formed NS's surface is still strongly magnetized. This can lead to a detectable radio precursor. 
	\end{itemize}
	
	The single most important conclusion of this work is that one should be open minded when interpreting short GRBs, BNS GW signals, $r$-process enriched stars and SNe in terms of the potential time delay between the events and the formation of the system. Specifically, a few hundred short GRBs have been detected so far. Among those, a few must have been ultrafast mergers with merger times of $\lesssim 30$\,Myr. Identifying them poses an interesting challenge. 
	
	\bigskip
	\noindent {\bf ACKNOWLEDGEMENTS}
	
	\medskip
	This research was supported by a grant (no. 2020747) from the United States-Israel Binational Science Foundation (BSF), Jerusalem, Israel and by a grant (no. 1649/23) from the Israel Science Foundation [PB], by an advanced ERC grant and by the Simons Collaboration on Extreme Electrodynamics of Compact
	Sources (SCEECS)  [TP]. 
	\medskip
	
	\noindent {\bf DATA AVAILABILITY}
	
	\medskip
	No data was generated by this work.

\end{document}